\documentclass{article}
\usepackage{bookstyle}
\usepackage{graphicx}
\usepackage{amsmath,amssymb}
\usepackage{cite}
\def\gsim{\mbox{\raisebox{-.6ex}{~$\stackrel{>}{\sim}$~}}}
\def\lsim{\mbox{\raisebox{-.6ex}{~$\stackrel{<}{\sim}$~}}}
\def\dsl{\hbox{\phantom{.}/\kern-.4600em$\partial$}}
\def\qsl{\hbox{\phantom{.}/\kern-.4600em$q$}}
\def\psl{\hbox{\phantom{.}/\kern-.4600em$p$}}
\def\beq{\begin{equation}}
\def\eeq{\end{equation}}
\def\beqa{\begin{eqnarray}}
\def\eeqa{\end{eqnarray}}
\begin{document}
\author{James M.\ Cline}
\address{McGill University, Dept.\ of Physics, Montreal, Qc H3A 2T8,
Canada}
\title{String Cosmology}
\photo{\includegraphics{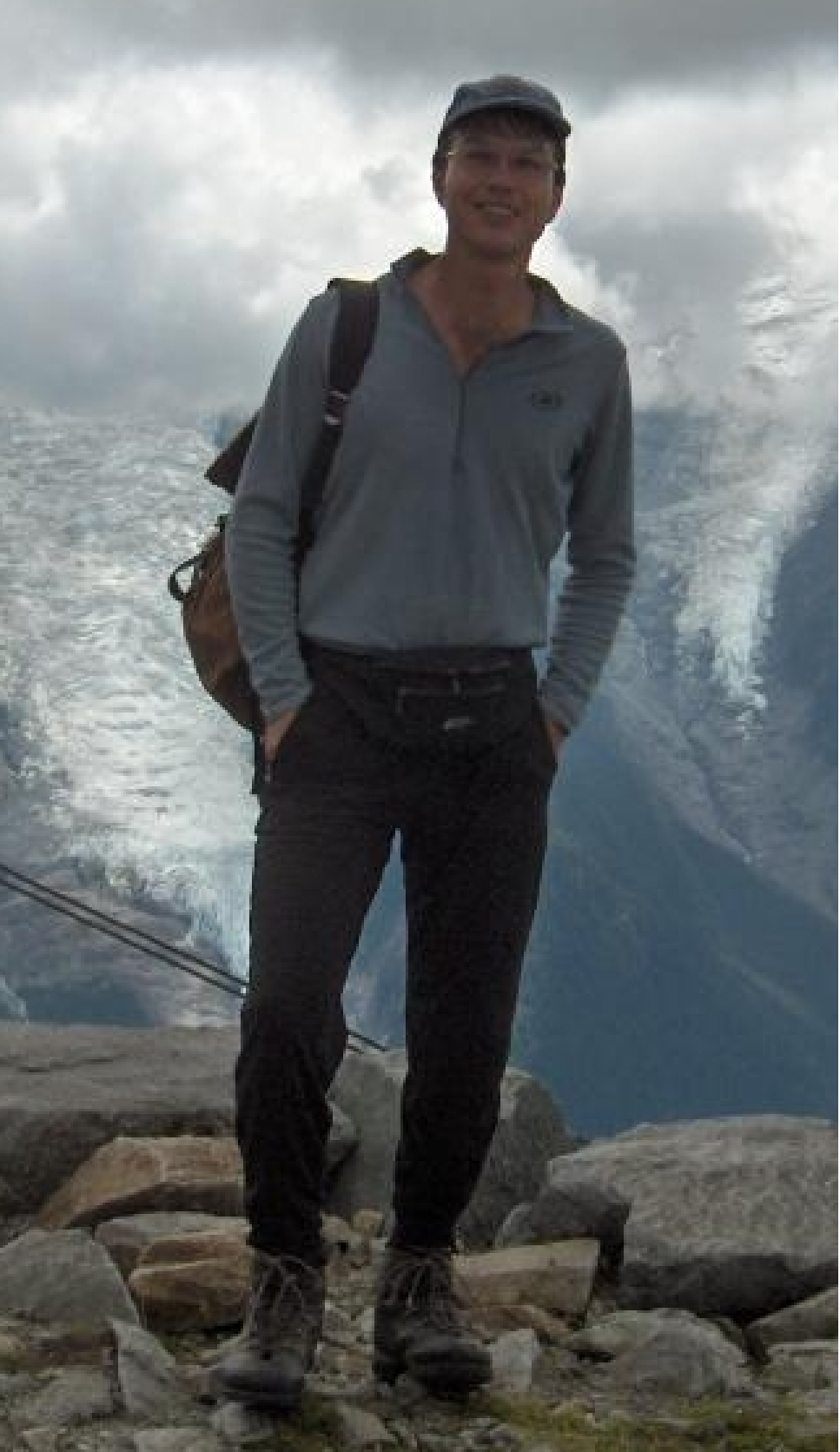}}
\frontmatter
\maketitle    
\mainmatter
\setcounter{figure}{1}

\section{Dark Energy}

\subsection{The problem of scales}
\label{sect1.1}
In preparing a set of lectures on string cosmology, I originally set
out to explain why I am going to focus exclusively on inflation from
string theory, ignoring the important question of the dark energy of
the universe, which is another very exciting area on the observational
side.  The basic reason is that the string scale is so much greater
than the scale of the dark energy that it seems implausible that a
stringy description should be needed for any physics that is occuring
at the milli-eV scale.  

Naively, one can generate small energy scales from string-motivated
potentials; for example exponential potentials often arise,
so that one might claim a quintessence-like Lagrangian of the form
\beq
	{\cal L } = (\partial\phi)^2 - \Lambda e^{-\alpha\phi}
\eeq
could be derived from string theory, and the small scale of the
current vacuum energy could be a result of $\phi$ having rolled to 
large values.  The problem with this kind of argument is that it is
ruled out by 5th force constraints, unless $\phi$ happens to be
extremely weakly coupled to matter.  But in string theory, there is
no reason for any field to couple with a strength that is suppressed
compared to the gravitational coupling, so it seems difficult to get
around this problem.  The E\"ot-Wash experiment \cite{eotwash} bounds
the strength of this particle's coupling to matter as a function of
its inverse mass (the range of the 5th force which it mediates) as
sketched in figure 1.  The current limit restricts a fifth force
of gravitational strength to have a range less than $0.044$ mm,
corresponding to a mass greater than $4.4$ milli-eV.  On the other
hand a field that is still rolling today must have a mass less than
the current Hubble parameter, $\sim 10^{-33}$ eV.

\centerline{
\includegraphics[width=0.9\textwidth]{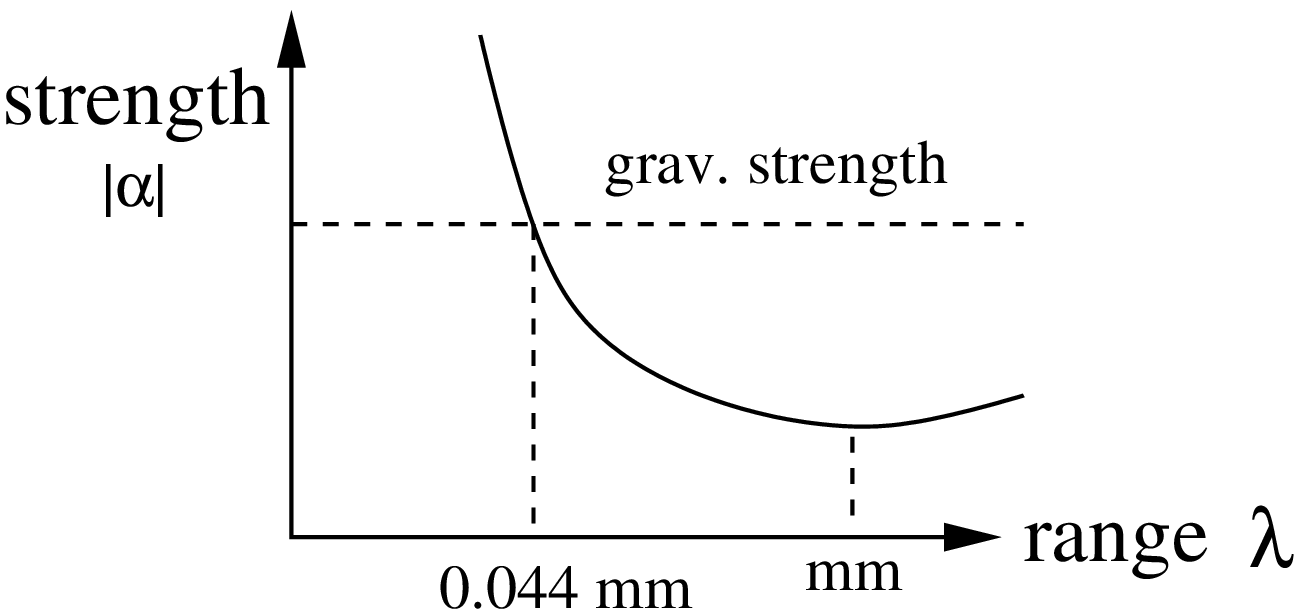}}
\centerline{Fig.\ 1. Schematic limit on coupling strength versus range of
a fifth force.}
\setcounter{figure}{1}
\bigskip

\subsection{The string theory landscape}

On the other hand, if we accept the simplest hypothesis, which is
also in good agreement with the current data, that the dark energy is
just a cosmological constant $\Lambda$, then string theory does have
something to tell us.  The vast landscape of string vacua
\cite{Susskind}, 
 combined
with the selection effect that we must be able to exist in a given
vacuum in order to observe it (the anthropic principle) has given the
only plausible explanation of the smallness of $\Lambda$ to date.

\begin{figure}[h]
\centerline{
\includegraphics[width=0.9\textwidth]{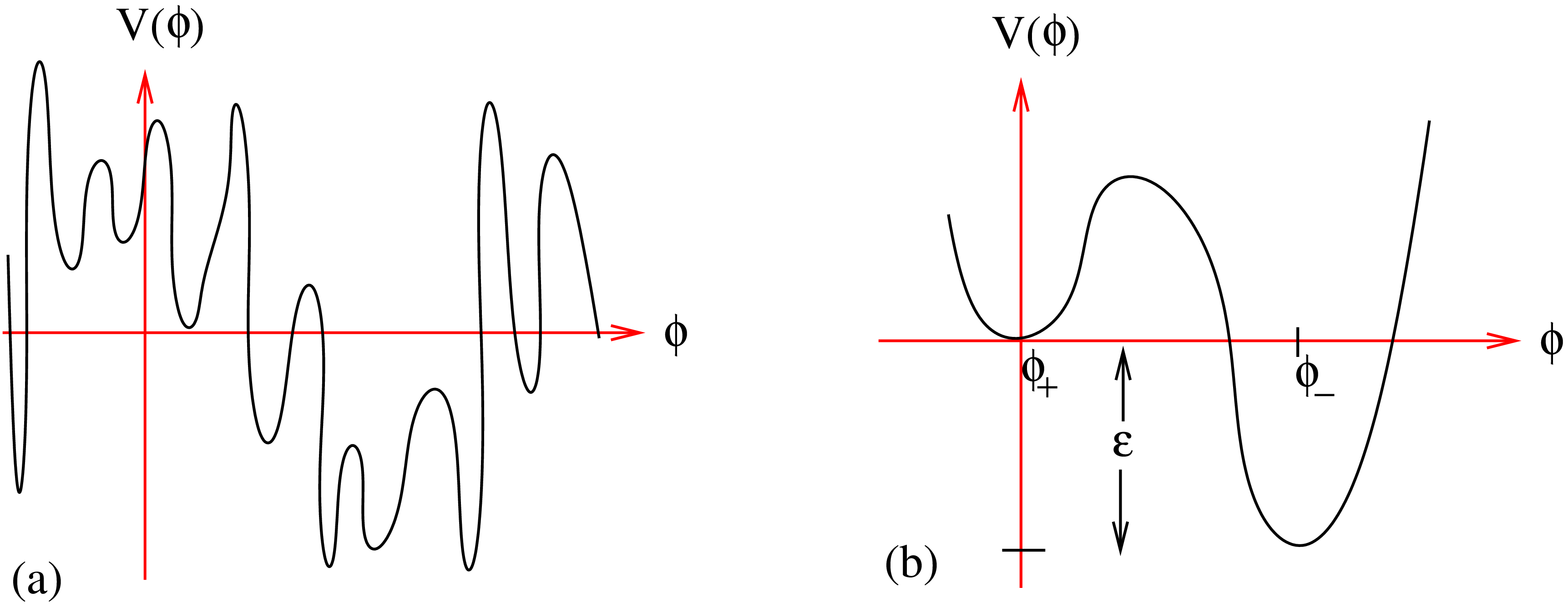}}
\caption{(a) Scalar field potential which is a toy model for 
the landscape.  (b) A part of the potential illustrating
the failure of tunneling, {\it \`a la} Coleman-DeLuccia.}
\label{fig2}
\end{figure}

We can illustrate the idea starting with a toy model, of a scalar
field with Lagrangian
\beq
	{\cal L } = (\partial\phi)^2 - V(\phi)
\eeq
whose potential has many minima, as shown in figure \ref{fig2}(a).
Naively one might guess that eventually the universe must tunnel
to the lowest minimum, which would generically be a negative energy
anti-de Sitter space ending in a catastrophic big crunch.  This is
wrong, for two reasons.  The first reason is that the lifetime of a 
metastable state may
be longer than the age of our universe, in which case it is as
good as stable.  More than that, a positive energy vacuum which would
be unstable to tunneling to a negative energy vacuum
in the absence of gravity can actually be stable when gravity is
taken into account.  As shown by Coleman and De Luccia \cite{CDL},
if the difference $\epsilon$ in vacuum energies between a zero-energy
minimum and one with negative energy is too great, tunneling is
inhibited.  Referring to figure \ref{fig2}(b), define 
\beq
	S_1 = \int_{\phi_-}^{\phi_+} d\phi
\sqrt{2(V_0(\phi)-V_0(\phi_+)}
\eeq
where, roughly speaking, $V_0$ is what $V$ would look like if
$\epsilon$ were set to zero.  This is well-defined when $\epsilon$
is small, and the bubbles of true vacuum which nucleate during the
tunneling consequently have a thin wall.  The initial radius of the
bubbles in this case turns out to be $\bar\rho_0 = 3 S_1/\epsilon$.
On the other hand, there is a distance scale $\Lambda =
\sqrt{3/(\kappa\epsilon)}$ (where $1/\kappa = 8\pi G$) which is the
size of a bubble whose radius equals its Schwarzschild radius.
Coleman and De Luccia show that if $\bar\rho_0 \ge 2\Lambda$, there is
no vacuum decay.  

The second reason to consider vacua with large energies is 
eternal inflation \cite{Linde}.  In regions where the potential
is large, so are the de Sitter quantum fluctuations of the field,
$\delta\phi\sim H/2\pi$.  If they exceed the distance by which the
field classically rolls down its potential during a Hubble time $1/H$,
then the quantum effects can keep the field away from its minimum
indefinitely, until some chance fluctuations send it in the direction
of the classical motion in a sufficiently homogeneous spatial region.
The classical motion during a Hubble time, $\Delta\phi$, 
can be estimated using 
the slow-roll approximation to the field equation, $3 H\dot\phi =
-V'$, so $\delta\phi = -V'/3H^2$.  Comparing this to the quantum
excursion, we see that the condition for eternal inflation is
\beq
	H^3 \gsim |V'|
\label{eternal}
\eeq
In this picture, the global universe consists of many regions
undergoing inflation for indefinitely long periods, occasionally
giving rise to regions in which inflation ends and a subuniverse like
ours can emerge.  Thus any minimum of the full $\phi$ potential which
is close to values of $\phi$ where (\ref{eternal}) can be fulfilled
will eventually be populated parts of the landscape.

\begin{figure}[h]
\centerline{
\includegraphics[width=0.7\textwidth]{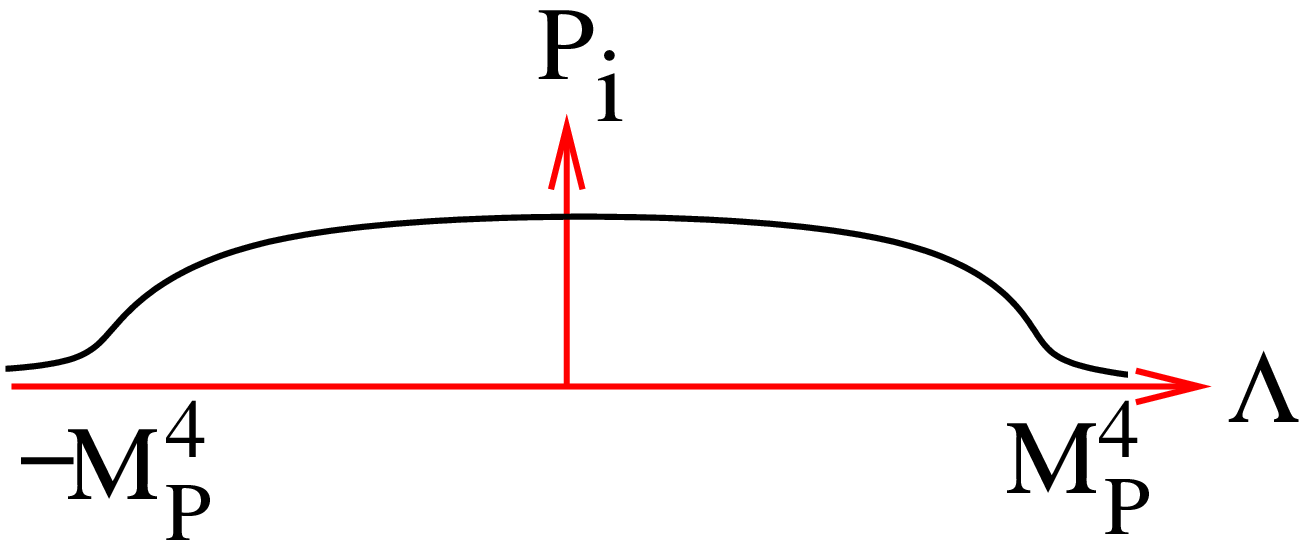}}
\caption{Intrinsic probability distribution for values of $\Lambda$
coming from minima of the landscape potential.}
\label{fig3}
\end{figure}

Now we must make contact with the cosmological constant, whose value 
locally depends on which of the many minima a given subuniverse
finds itself in.  Suppose there are $N$ which are accessible through
classical evolution or tunneling.  It is as good a guess as any to assume that the
values of the minima of the potential, $V_{\rm min}$ are uniformly
distributed over the range $\sim [-M_p^4, M_p^4 ]$ with spacing
$\Delta V \sim M_p^4/N$.  If $N\gsim 10^{120}$ then there will exist
some vacua with $V_{\rm min}$ close to the observed value.  The
possible values of $\Lambda$ will be distributed according to some
intrinsic probability distribution function  $P_i(\Lambda)$ which
depends on the details of the potential $V$, but in the absence of
reasons to the contrary should be roughly uniform, as in b
\ref{fig3}.
 
This theory explains why it is possible to find ourselves in a vacuum
with the observed value of $\Lambda$.  To understand why we were so
lucky as to find it so unnaturally close to zero, we need to consider
the conditional probability
\beq
	P_{\rm tot}(\Lambda) = P_i(\Lambda) P_{\rm obs}(\Lambda)
\eeq
where the second factor is the probability that an observer could
exist in a universe with the given value of $\Lambda$.  This part
of the problem was considered by S.\ Weinberg originally in \cite{WeinA}
in also in the review article
\cite{WeinR} .  In the abstract
of the \cite{WeinA}, he says that the anthropic bound on $\Lambda$
is too weak to explain the observed value of $\Lambda$, but in the
later work he took the more positive view; living in a universe whose
probability is 1\% is far less puzzling than one where $P\sim
10^{-120}$.  There are actually two anthropic bounds.  For
$\Lambda>0$, we must have $\Lambda\lsim 10^2\Lambda_{\rm obs}$;
otherwise the universe expands too quickly to have structure formation
by redshifts of $z\sim 4$.  For $\Lambda <0$, we need $\Lambda \gsim
-\Lambda_{\rm obs}$ to avoid the recollapse of the universe before
structure formation.  

It is worth pointing out that this was a {\it successful prediction}
of $\Lambda_{\rm obs}$ before it was determined to be  nonzero through
observations of distant supernovae \cite{SCP}.
From the anthropic point of view, there is no reason for $\Lambda$ to
be zero, so its most natural value is of the same order of magnitude
as the bound---assuming of course that $P_i$ is roughly uniform over
the range where $P_{\rm obs}$ is nonnegligible.

\subsection{The Bousso-Polchinski (Brown-Teitelboim) mechanism}

We can now consider how the setting for this idea can be achieved
within string theory.  The most concrete example is due to Bousso
and Polchinski \cite{BP}, who found a stringy realization of the
Brown-Teitelboim (BT) mechanism \cite{BT}.  This is based on the existence
of 4-form gauge field strengths $F^{(4)}_{\alpha\beta\gamma\delta}$
whose action is
\beq
	S = \int d^4x\left[ -\sqrt{-g} F^{(4)}_{\alpha\beta\gamma\delta}
	F_{(4)}^{\alpha\beta\gamma\delta}
	+ 8 \partial_\mu\left(\sqrt{-g}F_{(4)}^{\mu\alpha\beta\gamma}
	A_{\alpha\beta\gamma}\right)\right]
\label{4act}
\eeq
leading to the equation of motion  
$\partial_\mu\left(\sqrt{-g}F_{(4)}^{\mu\alpha\beta\gamma}\right)=0$,
with solution $F_{(4)}^{\mu\alpha\beta\gamma} = c 
\epsilon^{\mu\alpha\beta\gamma}$ where $c$ is a constant.
 The contribution to the action is
\beq
	S = + \int d^4x\sqrt{-g}\, (-4!)\, c^2
\label{4act2}
\eeq
The sign $+$ in front of the integral in (\ref{4act2}) would have
been $-$ had it not been for the total derivative term in 
(\ref{4act}) which must be there for consistency; otherwise the
value of the vacuum energy appearing in the action has the opposite
sign to that appearing in the equations of motion.  This negative
contribution to the action is a positive contribution to the vacuum
energy density.  

Although in field theory the constant $c$ is arbitrary, in string
theory it is quantized.  The particular setting used by Bousso
and Polchinski is M-theory, which is 11 dimensional and has no
elementary string excitations, but it does have M2 and M5 branes.
The low-energy effective action for gravity and the 4-form is
\beq
	S = 2\pi M_{11}^9 \int d^{11}x \sqrt{-g_{11}} \left(
	R - F_{(4)}^2\right)
\eeq
The 4-form is electrically sourced by the M2-branes and 
magnetically sourced by the M5-branes, in a way which we will 
discuss shortly.  To understand the quantization of $c$, it is 
useful to note that $F_{(4)}$ is Hodge-dual to a 7-form in 11
dimensions,
\beq
	F_{(4)} = * F_{(7)}
\label{dual}
\eeq
(in terms of indices, this means contracting $F_{(7)}$ with the
11D Levi-Civita tensor).  We can say that $F_{(7)}$ is electrically
sourced by M5 branes.  In the same way that electric charge in 4D
is quantized if magnetic monopoles exist, there is a generalized
Dirac quantization condition in higher dimensions which applies here
because the wave function of an M5 brane picks up a phase when
transported around an M2 brane, or {\it vice versa}.  For the wave
function to be single valued, it is necessary that the 
$F_{(7)}$ flux which is sourced by the M2 brane, when 
integrated over the 7D compactification
manifold $K$, is quantized,
\beq
	2\pi M_{11}^6 \int_K F_{(7)} = 2\pi n
\eeq
This quantization also implies the quantization of $F_{(4)}$ through
the duality relation (\ref{dual}).   With this identification, we
can evaluate the contribution to $\Lambda$ from (\ref{4act2}) as
\beq
	\Delta\Lambda = 2n {(2\pi M_{11}^3)^2\over 2\pi M_{11}^9 V_7}
	= {4\pi n\over M_{11}^3 V_7} \sim {n M_{11}^6\over M_p^2}
\eeq 
where $V_7$ is the volume of $K$.
This can be taken as the typical spacing between nearby values of 
the cosmological constant.  Unfortunately, even if $M_{11}$ is as
small as the TeV scale using large extra dimensions, one finds that
$\Delta\Lambda/M_p^4\sim ($TeV$/10^{15}$ TeV$)^6\sim 10^{-90}$, and
this spacing is too large to solve the cosmological constant problem,
which needs $10^{-120}$.  In fact the problem is even worse, because
if the bare value of $\Lambda$ which needs to be canceled is negative
and of order $-M_{11}^4$, a large value of $n$ is required,
\beqa
	\Lambda_{\rm obs} &=& \Lambda_{\rm bare} + O\left({n^2
M_{11}^6\over M_p^2}\right)\nonumber\\
	&\Longrightarrow& n\sim {M_p|\Lambda_{\rm bare} |^{1/2}\over
	M_{11}^3}
\eeqa
in which case $\Delta\Lambda/M_p^4\sim  M_{11}^3
	|\Lambda_{\rm bare} |^{1/2}/ M_p^5\sim 10^{-75}$
assuming $|\Lambda_{\rm bare} |\sim M_{11}^4$.

\begin{figure}[h]
\centerline{
\includegraphics[width=0.5\textwidth]{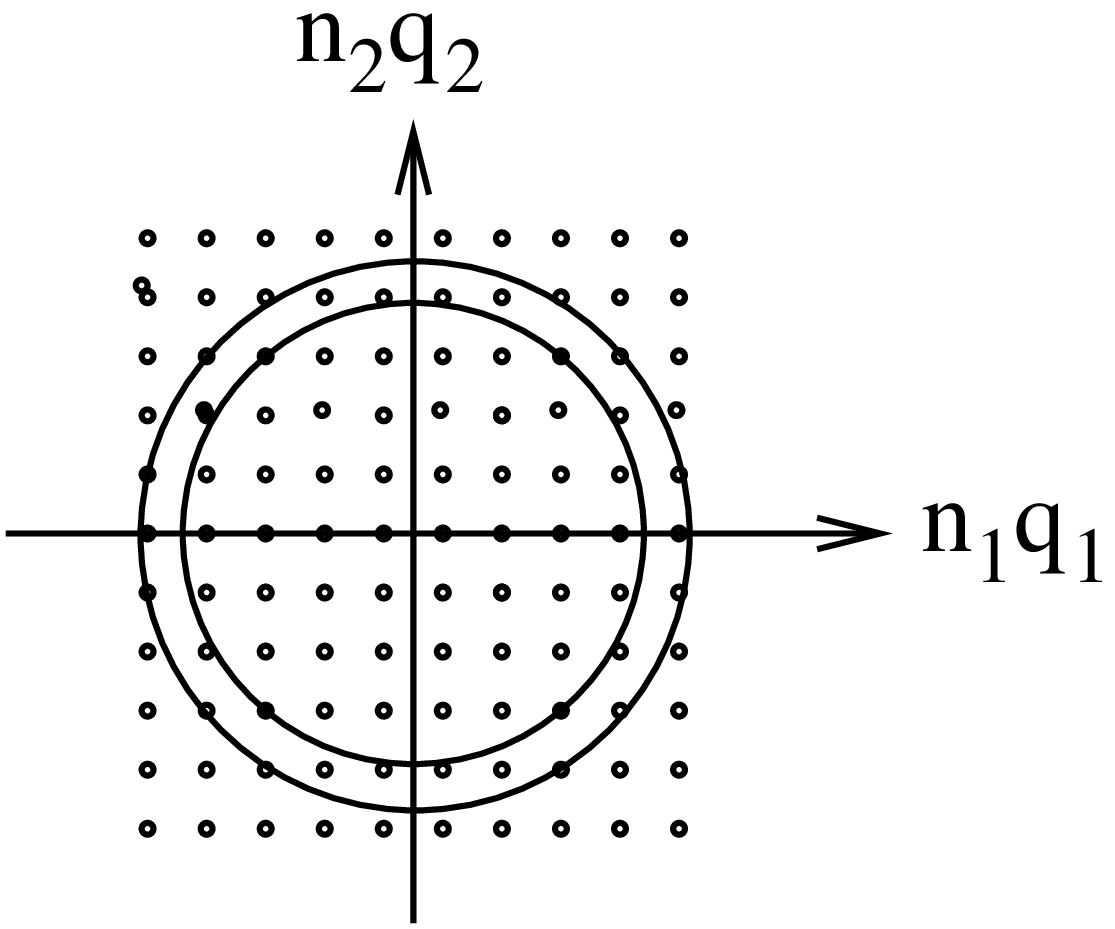}}
\caption{Lattice of allowed 4-form flux integers, and a hyperspherical
shell of values consistent with canceling a large negative
$\Lambda_{\rm bare}$ to get a small $\Lambda_{\rm obs}$.}
\label{fig4}
\end{figure}

However, string theory has a nice solution to this problem.  There is
not just one 4-form, but one for each nontrivial 3-cycle in $K$.
Then
\beq
	 F_{(7)} = *F_{(4,N+1)}\wedge \epsilon_1(y) +
	\sum_{i=1}^N F_{(4,i)}\wedge \omega_{3,i}(y)
\eeq
where $F_{(4,N+1)}$ can be thought of as the original 4-form we
started with, while $F_{(4,i)}$ is the $i$th additional one which 
arises for each nontrivial 3-cycle.  Here $\epsilon_1(y)$ is the
volume form on $K$ (whose coordinates are taken to be $y$) and 
$\omega_{3,i}(y)$ is the harmonic 3-form on the $i$th 3-cycle.
There can easily be a large number of 3-cycles.  For example if $K$ is
a 7-torus, the number of inequivalent triples of 1-cycles is
$({7\atop 3}) = 35$.  If $K$ has more interesting topology the number
can be larger.  We no longer have a single flux integer $n$ but a
high-dimensional lattice, $n_i$, such that
\beq
	\Lambda_{\rm obs} = \Lambda_{\rm bare} +\frac12 \sum_i n_i^2
	q_i^2
\eeq
where $q_i = M_{11}^{3/2} V_{3,i}/\sqrt{V_7}$ and $V_{3,i}$ is the 
volume of the $i$th 3-cycle.   We can visualize a spherical shell
in the space of the flux integers $n_i$, which is bounded by 
surfaces where $\Lambda = \Lambda_{\rm obs}$ and 
$\Lambda = \Lambda_{\rm obs}+\Delta\Lambda$, as illustrated in
figure \ref{fig4}.  As long as the lattice is fine enough so that at
least one point is contained within the shell, then there exists a set
of fluxes which can give a finely-enough tuned cancellation of
$\Lambda_{\rm bare}$.  The number of such lattice points is just the
volume of the shell with radius $\sqrt{\frac12\sum_i n_i^2 q_i^2}
= |\Lambda_{\rm bare}|^{\frac12}$ and
thickness $\Delta\Lambda^{\frac12}\sim \Lambda_{\rm obs}^{\frac12}$,
in the $N$ dimensional space whose $i$th axis is the magnitude of 
$n_i q_i$.  This fixes the magnitude of $q_i$ via
\beq
	{\Delta\Lambda\over M_p^4} \sim q_i^{N/2}\Gamma(N/2)\pi^{-N/2}
\label{dl}
\eeq

For example, if $N\sim 35$ we need  $q_i\sim 10^{-3}$ while if
$N\sim100$, we need $q_i\sim 10^{-2/3}$.  Let us suppose that the 
compactification and 3-cycle volumes are given in terms of a distance
scale $R$, as $V_7\sim R^7$, $V_{3,i}\sim R^3$.  Then
\beq
	q_i \sim {M_{11}^{3/2} R^{-1/2}\over M_p^2}
	\sim  {M_{11}^{3/2} R^{-1/2}\over M_{11}^{9} R^{7}}
	= {1\over (M_{11} R)^{13/2}}
\eeq
so the required small values of $q_i$ can be achieved by taking extra
dimensions which are moderately larger than the fundamental distance
scale.

\begin{figure}[h]
\centerline{
\includegraphics[width=0.5\textwidth]{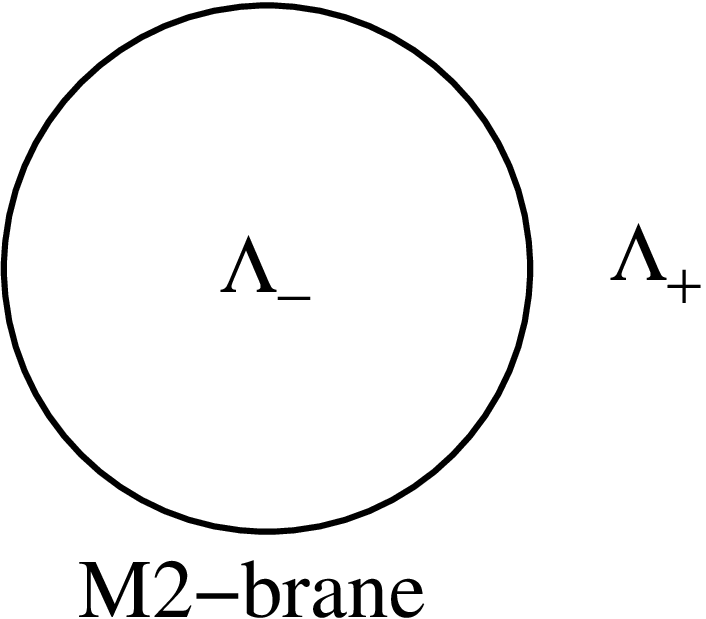}}
\caption{Nucleation of bubble of new $\Lambda$ whose wall is an
M2-brane.
}
\label{nuc}
\end{figure}

To complete the picture, we must describe how tunneling between the
different vacuum states (labeled by the flux quanta) takes place.
In the toy-model field-theory example, bubbles of the new vacuum phase
with $\Lambda_-$ nucleate within the old one with $\Lambda_+$.  In
M-theory there is no scalar field, but we do have M2 and M5 branes.
The M2 branes can serve as the walls of the bubbles since they are 
spatially two-dimensional, as illustrated in figure \ref{nuc}.
 Furthermore there is a natural way to
couple the M2 branes to the 4-forms, which is a generalization of the
coupling of a charged 
relativistic point particle to the Maxwell gauge field,
\beq
	S = -\int d\tau J^\mu A_\mu
\eeq
defined as an integral along the worldline of the particle,
where $\tau$ is the proper time and $J^\mu= q dx^\mu/d\tau$.  The
generalization to strings is
\beq
	S = -q \int d\tau d\sigma\, 
	{\partial x^\mu\over \partial\tau}\, 
	{\partial x^\nu\over \partial\sigma}\, A_{\mu\nu}
\label{CS1}
\eeq
and for 2-branes it is
\beq
	S = -q \int d\tau d\sigma_1 d\sigma_2 \, 
	{\partial x^\mu\over \partial\tau}\, 
	{\partial x^\nu\over \partial\sigma_1}\,
	{\partial x^\rho\over \partial\sigma_2}\, A_{\mu\nu\rho}
\label{CS2}
\eeq
where $A_{\mu\nu}$ and $A_{\mu\nu\rho}$ are totally antisymmetric
gauge potentials.  This shows why M2-branes are sources of 4-form
field strengths, since the latter are related to the 3-index gauge
potential by
\beq
	F_{(4)} = d A_{(3)}, \quad F_{\mu\nu\alpha\beta} = 
	A_{[\mu\nu\alpha,\beta]}
\eeq
(the brackets indicate total antisymmetrization on the indices).
Adding the Chern-Simons action (\ref{CS2}) to the kinetic term for
$F_{(4)}$ results in the equation of motion
\beq
	\partial_\mu(\sqrt{-g} F^{\mu\nu\alpha\beta}) = 
	q\sum_\mu \delta^{(1)}(x^\mu - X^\mu(\tau,\sigma_i))
	\epsilon^{\mu\nu\alpha\beta}
\eeq
For example if the M2 brane is in the $x$-$y$ plane
and the spacetime is Minkowskian, then
\beq
	\partial_\mu F^{\mu\nu\alpha\beta} = q\delta(z) 
	\epsilon^{z\nu\alpha\beta}
\eeq
whose solution is 
\beq
	F^{\mu\nu\alpha\beta} = \epsilon^{\mu\nu\alpha\beta}
	\left\{ {c+q,\ z>0\atop c,\qquad  z<0}\right.
\eeq
Thus the value of the 4-form changes by the charge of the M2 brane
when going from one side of the brane to the other.

In the BT mechanism, the value of $\Lambda$ decreases dynamically by
the nucleation of bubbles whose surface has  charge $q$
(figure \ref{nuc}).  We now
see why these bubble walls can be identified with the M2 branes
of M-theory, which carry the quantized charge $q$. Having
introduced a large number $N$ of 4-forms, we need to find separate
sources with different charges $q_i$, one for each different
$F_{(4,i)}$, yet M-theory only has one kind of M2 brane, with a
unique charge.  The source of the new $F_{(4,i)}$'s is the M5 brane,
whose extra three dimensions can be wrapped on the different
3-cycles.  Each one of these provides effectively a new kind of M2
brane.  The fact that these branes must come in integer multiples
also explains why the $F_{(4,i)}$ fluxes are quantized: since they
are sourced by M2 branes, their strength must be proportional to the
number of source branes.   We can now better understand  eq.\
(\ref{dl}): if we start from a situation with no fluxes  and nucleate
a stack of M2 branes, where $n_i$ copies of the $i$th kind of M2
brane are coincident, the new value of $F_{(4,i)}$ (after normalizing
it in a convenient way)  is given by $n_i q_i$, and the corresponding
contributions to $\Lambda$ add in quadrature.

This construction establishes that string theory contains the
necessary ingredients for realizing the original BT idea to explain
the smallness of $\Lambda$.  It is a concrete, detailed, and
quantitative treatment, which gives a clear explanantion of how many
closely-spaced values of $\Lambda$ can be generated.   It is a
nontrivial feat that it is possible to make the numbers work out
favorably.  There are also  further hurdles to pass: the desired
endpoint must have a lifetime greater than the age of the present
universe since it is only metastable---this can be arranged since
tunneling is typically exponentially suppressed---and the final
universe (ours) must be born in an excited state so that inflation
and reheating can take place subsequent to the nucleation.  The
latter requirement is less generic to satisfy, but possible, due to
eternal inflation.  If the inflaton is at an eternally inflating
value during the tunneling, the  daughter universe can also be
eternally inflating.  This puts a lower bound on $M_{11}$ around the
GUT scale.  

It is likely that other corners of the string theory parameter space can
also provide settings for this basic idea. The theory we will focus on for
inflation, type IIB string theory with flux compactification, gives
exponentially large ($10^{\sim 500}$) numbers of vacua \cite{Douglas}.  
(See \cite{dealwis} for a discussion of the BTBP mechanism in this
theory.)  

It is worth emphasizing that the use of the anthropic principle is
not merely optional once we accept the existence of many vacuum
states.  Unless there is a dynamical mechanism singling out one or a
few of the possible states, we are forced to consider all of them, 
and then to restrict attention to those which satisfy the prior that
observers can exist.  

\subsection{Caveats to the Landscape approach}

It has been noted that the anthropic bound on $\Lambda$ is relaxed if
the amplitude of density perturbations, $Q\sim 10^{-5}$ in our
universe, is also allowed to vary \cite{Q}:
\beq
	\hbox{bound on\ }\Lambda \sim Q^3
\eeq
Rees and Tegmark argue that if $Q>10^{-4}$, galaxies would be too
dense for life to evolve, which loosens Weinberg's bound by a factor
of $10^3$.  Nevertheless, a one part in $10^5$ fine-tuning is much
less daunting than one in $10^{120}$.   Furthermore, there could be
reasons for $Q$ not being ``scanned'' by the possible vacua of string
theory, when we understand them better.  

Another problem is the lack of additional predictions which would
allow us to test whether the anthropic explanation is really the
right one.  However, we are still in the early days of understanding
the landscape, so it may be premature to give up on making such
predictions.  Work along these lines is continuing at a steady pace
\cite{ADK,anthropic}.   One hope is that by studying large anthropically
allowed regions of the landscape, there might exist generic
predictions for other quantities correlated with a small value of
$\Lambda$, for example the scale of SUSY breaking. However the
predictions are not yet clear, with some authors arguing for a small
scale of SUSY breaking \cite{DGT} while others suggest a high scale,
for example the interesting possibility of ``split SUSY,'' in which
all the bosonic superpartners are inaccessibly heavy and only the
fermionic ones can be produced at low energies \cite{AHD,split}.  In such
scenarios the low Higgs mass would arise from the anthropic need for
fine tuning instead of any mechanism like SUSY for suppressing the
large loop contributions \cite{ADK}.  Unpalatable as some might find
such a possibility, it is nonetheless a distinctive prediction, which
if shown to be true would lend more weight to the landscape resolution
of the cosmological constant problem.

\section{Inflation}  String theory brings some qualitatively new
candidates for the inflaton, as well as some possible observable signatures
that are distinct from typical field theoretic inflation.  The
inflaton could be the separation between two branes within the compact
dimensions, or it could be moduli like axions associated with the
Calabi-Yau (C-Y) compactification manifold.  The new effects include large
nongaussianity and inflation without the slow-roll conditions being
satisfied, as in the DBI inflation scenario \cite{dcel,DBI}.  There are
also new issues connected with reheating in models with warped
throats.

\subsection{Brane-antibrane inflation}
D$p$-branes in type II string theory are dynamical $(p+1)$-dimensional
objects (with $p$ spatial dimensions) on which the ends of open
strings can be confined, hence giving Dirichlet boundary conditions to
the string coordinates transverse to the brane.  There has been
considerable progress in achieving stable compactifications
within type IIB string theory in the last few years, using warped
Klebanov-Strassler throats  with fluxes \cite{KS,GKP}.

To put these developments into perspective, let us recall some
differences between types of string theories.  Type I theory describes
both open and closed, unoriented strings with SO(32) gauge group.
Its low energy effective theory is that of $N=1$ supergravity.
Type II theories are of closed strings only (not counting the 
open string excitations which live on the D-branes themselves), with
$N=2$ SUGRA effective theories.  Type IIA theories contain D$p$-branes
with $p$ even, while IIB contains odd-dimensional branes.  Both of the
type II theories have the same action from the NS-NS sector---the
sector where the superpartners to the string coordinates $\psi^\mu$
have antiperiodic boundary conditions \cite{polbook}:
\beq
	S_{NS\hbox{-}NS} = {1\over \kappa_{10}^2}\int d^{\,10}x\,
	\sqrt{-g}\, e^{-2\Phi}\left( R + 4(\partial\Phi)^2 - 
	\frac12| H_3|^2\right)
\eeq
where $H_3$ is the 3-index Kalb-Ramond field strength, and $\Phi$
is the dilaton, which determines the string coupling via
$g_s = e^\Phi$.  The Ramond-Ramond (R-R) sector, where $\psi^\mu$
has periodic boundary conditions, exhibits the differences between
type IIA and IIB theories:
\beq
	S_{R\hbox{-}R} = - {1\over 4\kappa_{10}^2}\int d^{\,10}x\,
	\left\{\begin{array}{ll} |F_2|^2 + |F_4|^2,& IIA\\
	|F_1|^2 + |F_3|^2 + \frac12|F_5|^2,& IIB\end{array}\right.
\eeq
where $F_n$ is an $n$-index antisymmetric field strength.  These
expressions could have been written in a more symmetric way by
including higher values of $n$, and taking coefficients of $\frac12$
for each term, because of the duality relations
\beq
	F_6 = * F_4,\quad F_8 = * F_2,\quad F_5 = * F_5,\quad
	{\it etc.}
\eeq
$F_n$ couples to the D$p$-brane with $p=n-2$, as in eqs.\
(\ref{CS1}, \ref{CS2}).  Branes which have the wrong dimension 
for the theory in which they appear are unstable and quickly decay
into closed strings.

The Chern-Simons couplings of the branes to the gauge fields show that
branes are charged objects, and not simply delta-function sources of
stress-energy as is often assumed in phenomenological brane-world 
models.  This puts restrictions on the placement of branes within
compact spaces.  Since the gauge fields $F_n$ obey Gauss' law, 
the net charge of the branes in the compact volume must vanish.

\begin{figure}[h]
\centerline{
\includegraphics[width=0.5\textwidth]{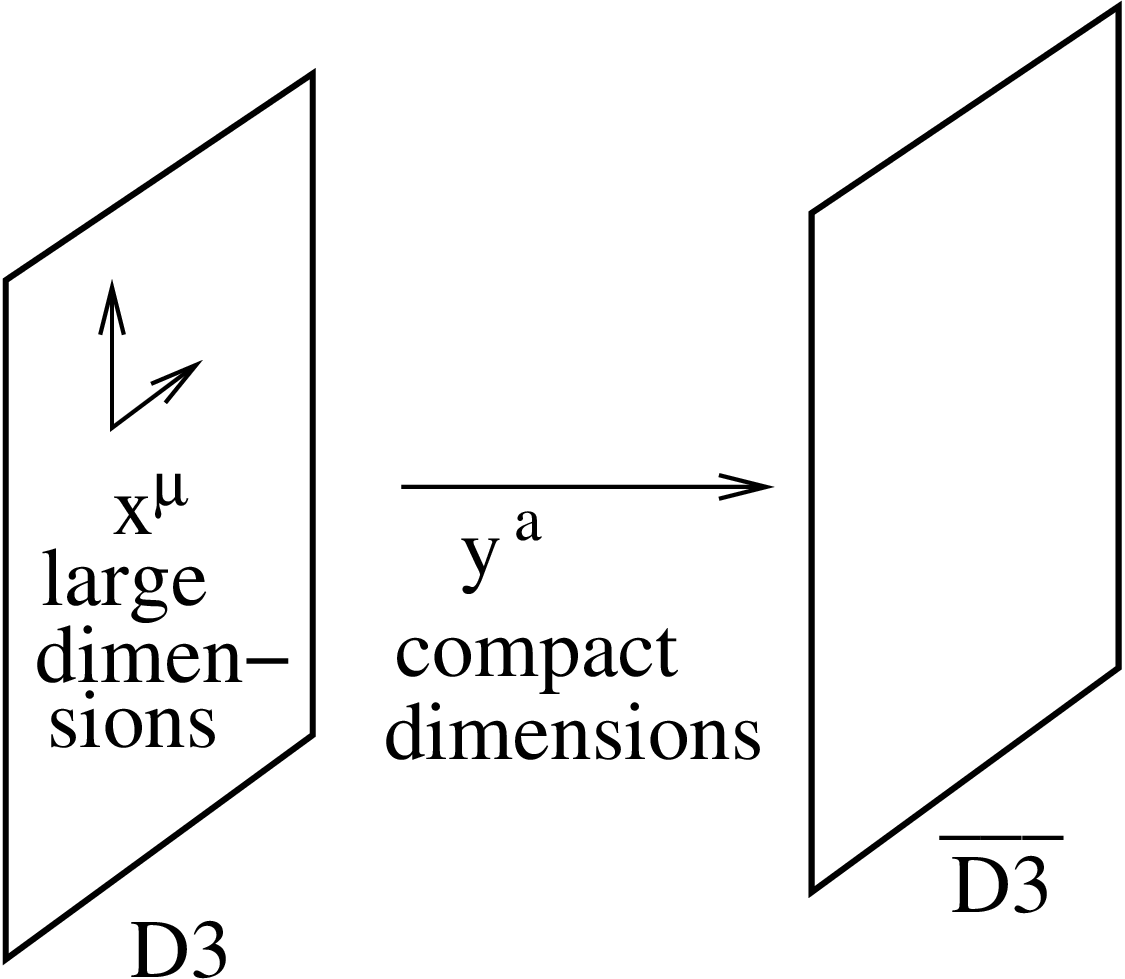}}
\caption{Parallel D3 brane and $\overline{\hbox{D3}}$ antibrane.
}
\label{bb}
\end{figure}

Consider for example a D3 brane parallel to a $\overline{\hbox{D3}}$
antibrane, which share parallel dimensions $x^\mu$ and are separated
by a vector $y^a$ in the transverse directions, lying in the
compact space.  This is illustrated in figure \ref{bb}.
They are sources of $F_5$ through their Chern-Simons
coupling, which can be more compactly written as
\beq
	\sum_i \mu_{D_3}^{(i)}\int_{M_4^{(i)}} A_4
\eeq
where $i$ labels the brane or antibrane having charge
$\mu_{D_3}^{(i)}$ and world volume $M_4^{(i)}$.   The equation of 
motion for $F_5$ is
\beq
	\partial_a\left(\sqrt{-g} F^{a\alpha\beta\gamma\delta}\right)
	+ \sum_i \mu_{D_3}^{(i)} \delta^{(6)}(y^a-y^a_i)
	\epsilon^{\alpha\beta\gamma\delta}
\eeq
where $a = 5,\dots,9$ and $\alpha\beta\gamma\delta = 0,1,2,3$
or some permutation.  Integrating this over the compact dimensions
gives
\beq
	\int d F + \sum_i \mu_{D_3}^{(i)}  = 0 
\eeq
Since the integral must vanish, so must the sum of the charges.

We can also use Gauss's law to find the force  associated with the
$F_5$ field, by considering noncompact extra dimensions, and 
integrating them over a 6D spherical region $R$ surrounding a single
D3 brane.  Then
\beq
	\int_R dF_5= \int_{\partial R} F_5 = |F_5|\times
\left(\hbox{area of 5-sphere}\right) \sum \mu_{D_3}
\eeq
This tells us that $|F_5|\sim \mu_{D_3}/r^5$ (the force) and 
$|A_4|\sim \mu_{D_3}/r^4$ (the potential), analogous to the $1/r$
Coulomb potential in 3D.  

Similarly, the gravitational potential for a D3 brane goes like 
$1/r^4$, so the total potential for a D3-D3 or 
D3-$\overline{\hbox{D3}}$ system with separation $r$ in the
compact dimensions is
\beq
	V_{\rm tot} = {G_{10}\over \frac14 \pi^2 r^4}\left(
	-\tau_3^2 \pm  {\mu_3^2\over g_s^2} \right)
	\left\{{\hbox{upper sign for D3-D3}\atop
	 \hbox{lower sign for D3-$\overline{\hbox{D3}}$}}\right.
\label{Vtot}
\eeq
where $\tau_3$ is the tension, generally given by 
\beq
	\tau_p = {M_s^{p+1}\over (2\pi)^p g_s}
\eeq
for a D$p$-brane, $M_s$ is the string mass scale, the 10D Newton 
constant is given by $G_{10} = (2\pi)^6 g_s^2/(8 M_s^8)$, and the
charge is related to the tension by $\mu_p = g_s\tau_p$.  Because of
this, the potential vanishes for D3-D3, but not for
D3-$\overline{\hbox{D3}}$.  We now consider whether this potential
can be used to drive inflation.

We would like to use the distance $r$ between D3 and 
$\overline{\hbox{D3}}$ as the  inflaton.  This indeed is a field,
since the branes are not perfectly rigid objects, but have transverse
fluctuations.  Hence at any longitudinal position $x^\mu$, the
distance between the brane and antibrane is
\beq
	r(x^\mu) = \left(\sum_a
	\left(y^a(x^\mu)- \bar y^a(x^\mu)\right)^2\right)^{1/2}
\eeq
We know the potential for $r$ from (\ref{Vtot}); all that remains is
to determine its kinetic term.  This is the Dirac-Born-Infeld (DBI)
action which for a D$p$-brane in Minkowski space is
\beq
	S = -\tau_p \int d^{\,p+1}x\, \left|
	\det\left(\eta_{\mu\nu} + \partial_\mu y^a\partial_\nu y^a
	\right)\right|^{1/2}
\eeq
in coordinates where $x^0,\dots,x^p$ are in the world-volume of the
brane and $a=p+1,\dots,9$ label the transverse coordinates.

For inflation we are interested in homogeneous backgrounds where 
$y^a = y^a(t)$ and so 
\beq
	\eta_{\mu\nu} + \partial_\mu y^a\partial_\nu y^a
	= {\rm diag}\left(-1 + \sum_a(\dot y^a)^2,\ +1,\ \dots,\ +1\right)
\eeq
Hence 
\beq
	|\det(\eta_{\mu\nu} + \partial_\mu y^a\partial_\nu y^a)|^{1/2}
	= \sqrt{1- \sum_a(\dot y^a)^2}
\eeq
In the limit of small velocities we can Taylor-expand the square root
to get the kinetic part of the action into the form
\beq
	S = \frac12 \tau_p \sum_a(\dot y^a)^2 + O(\dot y^4)
\eeq

\begin{figure}[h]
\centerline{
\includegraphics[width=0.5\textwidth]{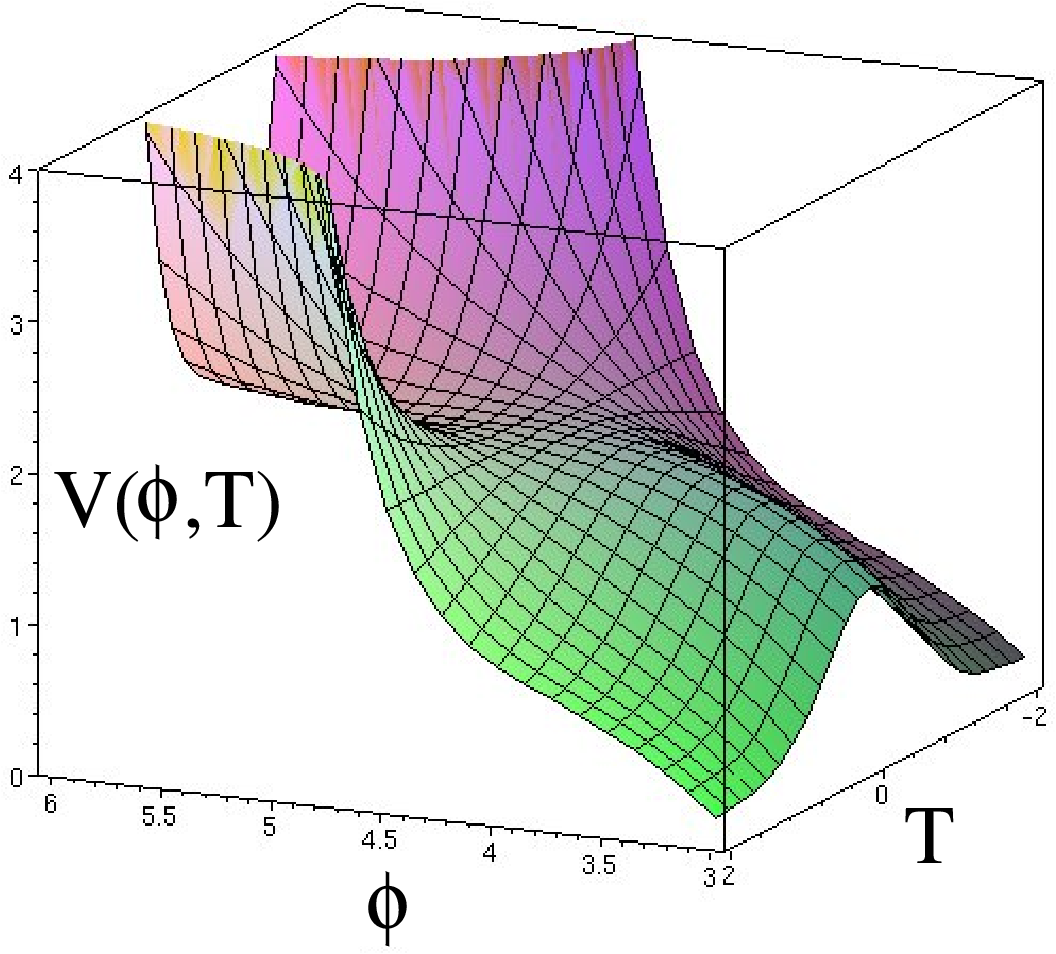}}
\caption{The hybrid-like potential for the brane separation $\phi$
and the tachyon $T$.}
\label{hybrid}
\end{figure}

It follows that the canonically normalized inflaton field is
\beq
	\phi = \sqrt{\tau_3}|\vec y - \vec{\bar y}|
\eeq
and the Lagrangian for D3 brane-antibrane inflation becomes
\beq
	{\cal L} = \frac12\dot\phi^2 + 2\left(\tau_3 - {c\over\phi^4}
	\right)
\eeq
where $c={4\over\pi^2} G_{10}\tau_3^4$.  This should not be taken
literally for $\phi^4< c/\tau_3$ however;  the Coulomb-like potential
is only a large-$r$ approximation.  The energy density never becomes
negative at small $r$.  A more careful calculation shows that $V$
remains finite as $r\to 0$, but there a tachyonic instability occurs
in another field when the brane-antibrane system reaches a critical
separation \cite{BS95}
\beq
	y_c^2 = {2\pi^2\over M_s^2}
\eeq
This is the separation at which the brane-antibrane system becomes
unstable to annihilation into closed strings.  The tachyon can be
seen as the ground state of the string which stretches between the
D3 and the $\overline{\hbox{D3}}$, whose mass is
\beq
	m^2_T = M_s^2\left({y^2\over y_c^2}-1\right)
\eeq
The full potential for $\phi$ and the tachyonic field $T$ resembles
that of hybrid inflation, fig.\ \ref{hybrid}.

Unfortunately, $V(\phi)$ is not flat enough to be suitable for
inflation, unless the separation $r$ is greater than the size of the
extra dimensions \cite{DT}, an impossible requirement \cite{Burgess}.
To see how this problem comes about, we compute the slow-roll
parameters
\beq
	\epsilon = \frac12 M_p^2\left({V'\over V}\right)^2
	\cong \frac12 M_p^2 \left( {4c\over \tau_3\phi^5}\right)^2
\eeq
\beq
	\eta = M_p^2\, {V''\over V} \cong - M_p^2\, {20 c\over \tau_3
\phi^6}
\eeq
where $M_p = 2.4\times 10^{18}$ GeV $= (8\pi G_N)^{-1/2}$.  What is
the relation between $G_N$ and $G_{10}$?  This can be found by
integrating out the extra dimensions from the 10D SUGRA action:
\beq
	S_{NS\hbox{-}NS} = {1\over 2\kappa_{10}^2 g_s^2}
	\int d^{\, 10}x\, \sqrt{-g_{10}} R_{10}
\implies {1\over 16\pi G_N} \int d^{\, 4}x\sqrt{-g} R
\eeq
The coefficient of the 10D action can also be written as $(16\pi
G_{10})^{-1}$, and after integrating over the extra dimensions,
assumed to have volume $L^6$, the 10D integral becomes $L^6\int d^{\, 4}x
\sqrt{-g} R$. Hence
\beq
	{L^6\over G_{10}} = {1\over G_N}
\eeq
and
\beq
	c = {4\over \pi^2} L^6 G_N \tau_3^4 = 
	{L^6 \tau_3^4\over 2\pi^2 M_p^2}
\eeq
This allows us to evaluate the slow-roll parameters as
\beq
	\epsilon \sim {M_s^{24} L^{12}\over M_p^2\phi^{10}},\quad
	\eta \sim {M_s L^6\over \phi^6}
\eeq

To get enough inflation, and to get the observed spectral index
\beq
	n_s = 2\eta-6\epsilon= 0.95\pm 0.02
\eeq
as observed by WMAP \cite{wmap3}, it is necessary to have small values
of $\epsilon$ and $\eta$.  The condition $\eta<1$ is the more
restrictive, and implies that the brane separation be
\beq
	r=y\sim {\phi \over M_s^2} \gg L
\eeq
which is the impossible situation of the branes being farther from
each other than the size of the extra dimensions \cite{Burgess}.
One can show that asymmetric compactifications do not help the
situation.  There were a number of attempts to solve this problem,
but these could not be considered complete because of the additional
unsolved problem of how to stabilize the moduli of the
compactification.  The early papers had to assume that the extra
dimensions were stabilized somehow, in a way that would not interfere
with efforts to keep the inflaton potential flat.  However, this is a
big assumption, as it later proved.

\subsection{Warped Compactification}
The problem of compactification was advanced significantly by
Giddings, Kachru and Polchinski (GKP) \cite{GKP}, building on work of
Klebanov and Strassler (KS) \cite{KS}.  The KS construction involved
putting branes at a conifold singularity, in order to reduce the number
of supersymmetries of the effective action on the D3-brane
world volume from $N=4$ 
down to the more realistic case of
$N=1$.  We need to understand this construction before studying its
applications to inflationary model building.

\begin{figure}[h]
\centerline{
\includegraphics[width=0.5\textwidth]{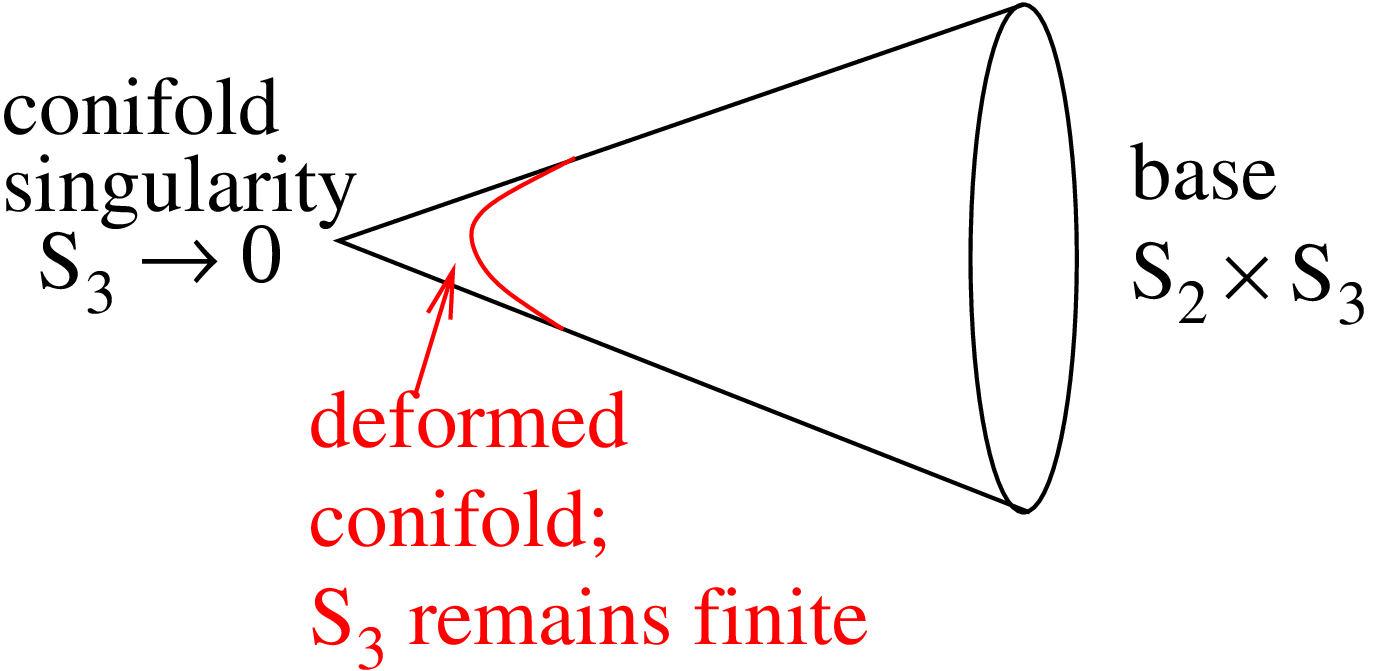}}
\caption{The conifold and deformed conifold.}
\label{cone}
\end{figure}

The conifold is a 6D Calabi-Yau space, which can be described in terms
of 4 complex coordinates $w_i$, constrained by the complex condition
\beq
	\sum_{i=1}^4 w_i^2 = 0
\label{cond}
\eeq
This looks similar to a cone, fig.\ \ref{cone}
and the singularity is at the tip, where
$w_i=0$.  The base of the cone has the topology of $S_2\times S_3$, 
a 5D manifold.  As one approaches the tip, the $S_3$ shrinks to a
point, and the topology of the tip is the remaining $S_2$.  The $S_3$
subspace is a 3-cycle, which is referred to as the A-cycle.  There
is also another (dual) 3-cycle, namely the $S_2$ times a
circle which is extended along the radial direction, and which we call
the B-cycle.   The whole 6D manifold times 4D Minkowski space
is a solution to Einstein's equations in 10D, and so it is a suitable
background for string theory.

It is possible to deform the conifold so that it is no longer
singular, by taking a more general condition than (\ref{cond}),
\beq
	\sum_{i=1}^4 w_i^2 = z
\eeq
Here $z$ is known as the complex structure modulus.  For $z\neq 0$,
the tip of the deformed conifold becomes a smooth point, at which the
$S_3$ is no longer singular but has a size determined by $z$.  The
deformed conifold becomes the solution to Einstein's equations when
certain gauge fields of string theory are given nonzero background
values.  These are the RR field $F_{(3)}$ of type IIB theory, and the
NS-NS Kalb-Ramond field $H_{(3)}$.  Since they are 3-forms, they can 
have nonvanishing values when their indices are aligned with some of 
the 3-cycles mentioned above, similar to turning on an electric field
along the 1-cycle (a circle) in the Schwinger model (electrodynamics in 
1+1 dimensions).  The lines of flux circulate and so obey Gauss's law
in this way.  In a similar way to the 4-form flux of heterotic
M-theory discussed in the previous section, these fluxes are also
quantized, due to the generalized Dirac quantization argument.
There are integers $M$ and $K$ such that
\beq
	{1\over 2\pi\alpha'}\int_A F_3 = 2\pi M,\qquad
	{1\over 2\pi\alpha'}\int_B H_3 = -2\pi K 
\eeq
where the slope parameter $\alpha'$ is related to the string mass
scale by $\alpha' = 1/M_s^2$, and A,B label the 3-cycles mentioned
above.  It can be shown that these 3-cycles are specified by the 
conditions $\sum_{i=1}^4 x_i^2 = z$  for the A-cycle and 
$x_4^2-\sum_{i=1}^3 y_i^2 = z$ if $w_n= x_n+i y_n$. The size of the
$S_3$ at the tip of the conifold gets stabilized by the presence
of the fluxes; the complex structure modulus takes the value
\beq
	z = e^{-2\pi K/(M g_s)} \equiv a^3(r_0)
\label{warpfactor}
\eeq
We will discuss the meaning of $a(r_0)$ shortly.

\begin{figure}[h]
\centerline{
\includegraphics[width=0.5\textwidth]{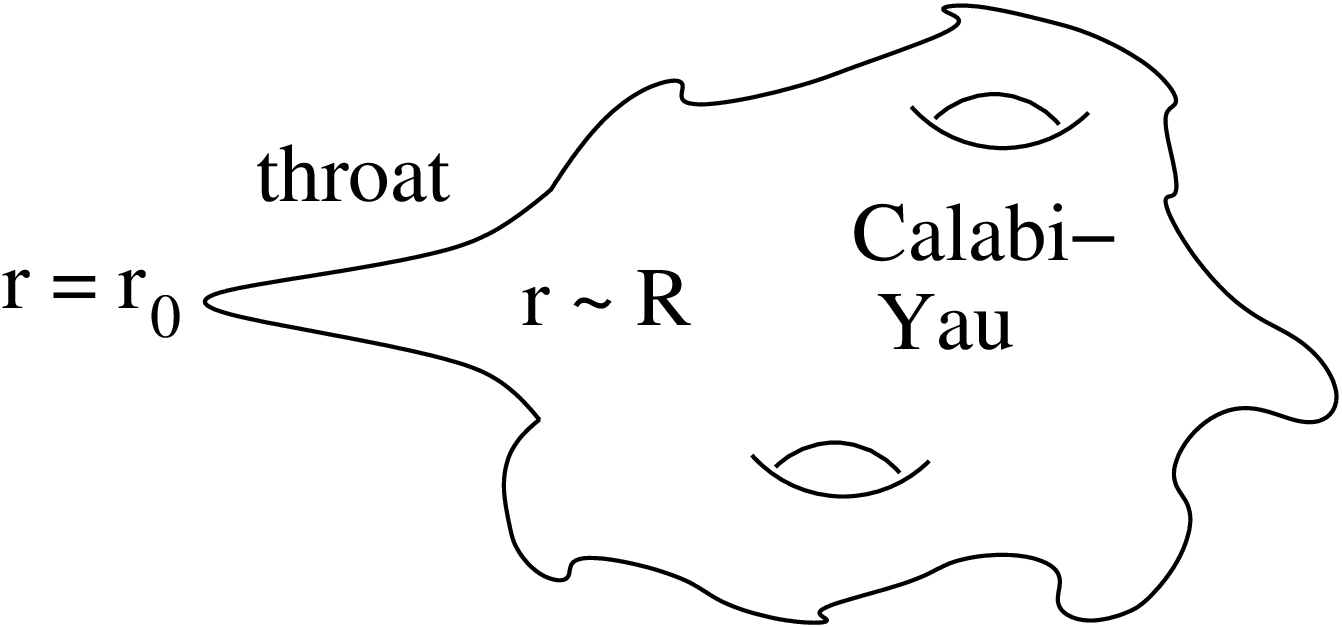}}
\caption{The KS throat glued to the larger Calabi-Yau manifold where the
radial coordinate $r\sim R$.}
\label{cy}
\end{figure}

When the fluxes are turned on, they also warp the geometry of the 
conifold.  The line element for the full 10D geometry is
\beq
	ds^2 = {dx^\mu dx_\mu\over \sqrt{h(r)}} + \sqrt{h(r)}
	\left(dr^2 + r^2 ds^2_{T_{1,1}}\right)
\label{KS}
\eeq
which is approximately AdS$_5\times T_{1,1}$.  The factor $T_{1,1}$
is the 5D base of the cone, known as an Einstein-Sasaki space, which
for our purposes is some compact angular space whose details will not
be important.  The warp factor $h(r)$ is approximately of the form
\beqa
	h(r) &=& {R^4\over r^4}\left( 1 + g_s {M\over K}\times
	(\ln r \hbox{\ correction} ) \right)\nonumber\\
	&\cong& {R^4\over r^4},\qquad R^4 = {27\over 4}\pi g_s
	N\alpha'^2,\quad N= MK
\eeqa
This is a stringy realization of the Randall-Sundrum (RS) model \cite{RS},
 known as
the Klebanov-Strassler throat, where the bottom of the throat (the tip
of the conifold) is at $r=r_0$, such that
\beq
	{r_0\over R} = z^{1/3} = a(r_0)= \hbox{\ warp factor}
\eeq
This explains the introduction of $a^3(r_0)$ in (\ref{warpfactor}).
The approximation $h=(R/r)^4$ gives the simple AdS$_5$  geometry 
for 4D Minkowski space 
times radial direction, and $R$ is the curvature scale of the
AdS$_5$.  The AdS part of the metric (\ref{KS}) can be converted
to the RS form $ds^2 = e^{\pm 2ky} dx^2 + dy^2$ through the change of 
variables $dy = \mp (R/r)dr$, $y = \mp R \ln r$, $r= e^{\mp ky}$,
where $k=1/R$.
As shown in fig.\ \ref{cy},
the top of the throat is understood to be smoothly glued onto the bulk
of the greater Calabi-Yau manifold, 
 whose geometry could be quite
different from that of (\ref{KS}). The gluing is done at some radius
$r\sim R$ where $h(r)\sim 1$.

Let us take stock of what the fluxes have done for us before we
continue with the search for inflation.  First, they stabilized the
complex structure modulus $z$, which without the fluxes was a
massless field, with an undetermined value.  In a similar way, they
stabilize the dilaton, which is essential for any realistic string
model since it determines the string coupling, and would give a ruled-out
5th force if massless.   Second, they have given us
warping, which introduces interesting possibilities for generating 
hierarchies of scales and generally having another parameter to tune.

Now we consider how the KS throat can be relevant for brane-antibrane
inflation.  We will see that a D3 brane by itself feels no force
in the throat, whereas the $\overline{\hbox{D3}}$ antibrane sinks to
the bottom.  This is because of the combination of gravitational and
gauge forces---they cancel for D3 but add for $\overline{\hbox{D3}}$.
To understand how this comes about, recall that the D3 brane is the
source of the $F_{(5)}$ field strength, which comes from the
$C_{(4)}$ gauge potential.  The fluxes create a background of 
$C_{(4)}$ or $F_{(5)}$ because there is contribution to the $F_{(5)}$
equation of motion (which we did not previously mention) of the form
\beq
	dF_5 \sim H_3 \wedge F_3
\eeq
Thus the complete action for a brane or an antibrane 
located at $r = r_1(x^\mu)$ is 
\beq
	S = -\tau_3\int d^{\,4}x {1\over h(r_1)}
	\sqrt{1 - h(r_1)(\partial r_1)^2}
	\pm \tau_3 \int d^{\,4}x (C_4)_{0123}
\label{bs}
\eeq
where the sign is $+$ for D3 and $-$ for $\overline{\hbox{D3}}$.
The form of the DBI part of the action can be understood from the
more general definition,
\beq
	S_{DBI} = -\tau_3\int d^{\,4}x \sqrt{-G}
\eeq
where the induced metric on the D3-brane is given by
\beq
	G_{\mu\nu} = G_{AB}{ \partial X^A\over \partial x^\mu}
	{ \partial X^B\over \partial x^\nu} = 
	{1\over \sqrt{h}}\eta_{\mu\nu} -\sqrt{h}\, \partial_\mu r\,
	\partial_\nu r
\eeq
in the case where $r$ is the only one of the 6 extra dimensions which
depends on $x^\mu$.  Furthermore the equation of motion for the RR
field has the solution
\beq
	(C_4)_{\alpha\beta\gamma\delta} = {1\over h(r_1)}
	\epsilon_{\alpha\beta\gamma\delta}
\label{C4sol}
\eeq

Consider the case where the transverse fluctuations of the brane
vanish, $\partial r_1=0$.  Then the two contributions to the action 
cancel for D3, but they add for $\overline{\hbox{D3}}$, to give
\beq
	S = -2\tau_3 \left({r_1\over R}\right)^4
	\int d^{\,4}x = -2\tau_3 a^4 (r_1)\int d^{\,4}x
\eeq
Notice that $\tau_3  a^4$ is the warped brane tension, and $V=\tau_3
a^4$ is the 4D potential energy associated with this tension.  Because
of the warp factor, $V$ is minimized at the bottom of the throat.
This is why the antibrane sinks to the bottom of the throat, whereas
the D3 is neutrally buoyant---it will stay wherever one puts it.  

To be consistent, we should recall the words of warning issued earlier
in these lectures: since branes carry charge, one is not allowed to 
simply insert them at will into a compact space.  The background must
be adjusted to compensate any extra brane charges.  This results in
a {\it tadpole} condition
\beq
	{\chi\over 24} = N_{D3} - N_{\overline{{D3}}}
		+ {1\over \kappa_{10}^2 T_3}
	\int_{C-Y} H_3\wedge F_3
\label{tadpole}
\eeq
where $\chi$ is the Euler number of the C-Y, and $N_X$ is the number of
branes of type $X$.  This relation says that
for a fixed topology of the C-Y, any change in the net D3 brane charge
has to be compensated by a corresponding change in the fluxes.
The flux contribution is positive, so there is a limit on the net
brane charge which can be accommodated, and this limit is determined
by the Euler number, which can be as large as $\sim 10^4$ in some
currently known C-Y's.  

\subsection{Warped brane-antibrane inflation}

With this background we are ready to think about brane-antibrane
inflation in the KS throat, following KKLMMT \cite{KKLMMT}.  Now we want to add not
just a D3 or $\overline{\hbox{D3}}$ to the throat, but both
together.  To find the interaction energy requires a little more work
relative to the calculation of (\ref{bs}).  Following \cite{KKLMMT},
we will consider how the presence of a D3 at $r_1$ perturbs the
background geometry and $C_4$ field.  Using the perturbed background
in the action for the $\overline{\hbox{D3}}$ will then reveal the
interaction energy.  We can write the perturbation to the background
as
\beqa
	h(r) &\to& h(r) + \delta h(r)\\
	C_4(r) &\to& C_4(r) + \delta C_4(r)
\eeqa
where $\delta h$ is determined by the Poisson equation in 6D,
\beq
	\nabla^2 \delta h = C \delta^{(6)}(\vec r - \vec r_1)
\eeq
in analogy to the gravitational potential for a point mass in 4D;
$C$ is a constant which turns out to have the value
$C = R^4/N$ in terms of the AdS curvature scale and the product of
flux quantum numbers.  The solution is 
\beq
	\delta h = {R^4\over N r_1^4},\quad
	\delta C_4 = - {\delta h\over h^2}
\eeq
(see eq.\ (\ref{C4sol})).  Using the perturbed background fields, we can
evaluate the Lagrangian for an antibrane at the bottom of the throat
$r=r_0$,   and expand to leading order in $(\partial r_1)^2$, to 
find the result
\beqa
	{\cal L} &=& \frac12\tau_3(\partial r_1)^2 - 2\tau_3
	\left({r_0\over R}\right)^4\left( 1 - \left({r_0\over R}\right)^4
	\, {R^4\over N r_1^4}\right)\\
	&=& \frac12(\partial \phi)^2 - 2\tau_3 a_0^4\left(
	1- a_0^4 {R^4\tau_3^2\over N \phi^4}\right)
\eeqa	
where we introduced the canonically normalized inflaton
$\phi=\sqrt{\tau_3}r_1$ and $a_0 = a(r_0)$.

When we compute the slow roll parameters in the new theory
with warping, and compare them to our previous calculation without
the warp factors, we find that
\beqa
	\epsilon = a_0^8\, \epsilon_{\rm unwarped}\nonumber\\
	\eta = a_0^4\, \eta_{\rm unwarped}
\eeqa
Previously we had the problem that $\eta\sim 1$ unless the branes were
separated by more than the size of the extra dimensions.  With the
new powers of $a_0$, we can easily make $\epsilon\ll\eta\ll 1$ since
naturally $a\ll 1$.  It looks like warping has beautifully solved the
problem of obtaining slow roll in brane-antibrane inflation.

\subsection{The $\eta$ problem}

The story is that KKLLMMT had happily arrived at this
result, but then realized the following problem.
 In fact $\eta$ is not small because one has neglected the
effect of stabilizing the K\"ahler modulus (the overall size of the
Calabi-Yau), $T$.  The inflaton couples to $T$ in such a way that
when a mass is given to $T$, it also  inevitably gives a
new source of curvature to the inflaton potential which makes $\eta$
of order unity.  This is the well-known $\eta$ problem of
supergravity inflation models.

To appreciate this, it is useful to consider the low-energy effective
SUGRA Lagrangian for the K\"ahler modulus,
\beqa
	{\cal L} &=& G_{T\overline T} \partial_\mu T\partial^\mu\overline T
	- e^K\left(G^{T\overline T}\, D_T W \overline{D_T W} - 3|W|^2\right)\nonumber\\
	&\equiv& G_{T\overline T} \partial_\mu T\partial^\mu\overline T
- V_F
\eeqa
where $T=\rho + i\chi$ is the complex K\"ahler modulus, related to
the size $L$ of the Calabi-Yau through
\beq
	\rho \equiv e^{4u} \sim L^4,
\eeq
$\chi$ is the associated axion, 
\beq
	G_{T\overline T}  = \partial_T\partial_{\overline T} K 
\eeq
is the K\"ahler metric, 
\beq
	K = -3\ln(T+\overline T)
\eeq
is the  K\"ahler potential, $W(T)$ is the superpotential, 
\beq
	D_T W = \partial_T W + W\partial_T K
\eeq
is the covariant derivative of $W$, and $V_F$ is the F-term potential.

In the GKP-KS flux compactification, $T$ remains unstabilized because
the fluxes generate a superpotential which only depends on the dilaton
and complex structure modulus, not on $T$.  Once these heavy fields are
integrated out, $W$ can be treated as a constant, $W_0$.  Then
\beqa
	D_T W &=& W\partial_T K = -3\, {W\over T + \overline T},\\
	G_{T\overline T} &=& {3\over ( T + \overline T)^2},\quad
	G^{T\overline T} = {( T + \overline T)^2\over 3},\\
	V_F &=& 3|W|^2- 3|W|^2 = 0
\eeqa
A SUGRA theory where this kind of cancellation occurs is called a
{\it no-scale} model.  

Of course having a massless K\"ahler modulus is unacceptable, for all
the reasons mentioned in section \ref{sect1.1}.  We need a nontrivial
superpotential $W(T)$.  We will come back to this.  First we focus on
the coupling which the inflaton---the position of the mobile
D3-brane---has to $T$, which will lead to
problems when we introduce $W(T)$.  Since K\"ahler manifolds are
complex, we can regard the position of the D3-brane in the 6 extra
dimensions as being specified by 3 complex coordinates, which become
three complex scalar fields, 
\beq
	\phi^i(x^\mu),\quad i=1,2,3
\label{infi}
\eeq
Previously we ignored the dependence on the angular directions in 
$T_{1,1}$ and only kept track of the radial position of the brane in
the throat, $r$, but (\ref{infi}) is the more precise specification.
It can be shown by several arguments (of which we will give one
shortly) that when there is a D3 brane 
in the throat, it changes the K\"ahler potential by
\beq
	K \to K\left(T+\overline T - k(\phi^i,\bar\phi^i)\right)
	\equiv K(2\sigma)
\label{Ksubs}
\eeq
where $k(\phi^i,\bar\phi^i)$ is a real-valued function, known as the 
K\"ahler potential for the Calabi-Yau (not to be confused with $K$
which is the K\"ahler potential for the field space).
 Now it is
the combination $2\sigma$ and not $T+\overline T$ which is related to
the physical size of the Calabi-Yau,
\beq
	2\sigma = L^4 = e^{4u} = V^{2/3}
\eeq
where $V$ is the volume of the extra dimensions.  It can be shown
(see for example \cite{HKN}) that
\beq
	k = \sum_i \phi^i\bar\phi^i + O(\phi^4)
\eeq
in the vicinity of the bottom of the throat, labeled by $\phi^i=0$.
(KKLMMT did not make this identification 
of $\phi^i=0$ with the bottom of the throat.)  

Next we can generalize the SUGRA action to take account of the
additional brane moduli fields, writing
\beq
{\cal L} = G_{I\overline J} \partial_\mu \Phi^I
\partial^\mu\overline\Phi^J
- e^K\left(G^{I\overline J} D_I W \overline{D_J W} - 3|W|^2\right)
\eeq
where $\Phi^I = (T,\phi^i)$.  One can show that the new definition of
$K$ still leads to a no-scale model if $W$ is constant, so $V_F$
remains zero in the presence of the D3 brane.  However the kinetic
term is modified; the K\"ahler metric for the brane moduli is
\beqa
	G_{i\bar\jmath} &=& {\partial^2K\over\partial\phi^i\partial
	\bar\phi^j} = {\partial\over\partial\bar\phi^j}
	\left({3\bar\phi^i\over T+ \overline{T} - k}\right)
	\nonumber\\
	&=& {3\delta_{ij}\over 2\sigma} + {3\phi^j\bar\phi^i\over
	2\sigma}\nonumber\\
\eeqa
where the last term is considered to be small because we want $\sigma$
to be large to justify integrating out the extra dimensions (so that
the corrections from higher dimension operators involving the
curvature are small), and this requires that $\phi^2\ll
T+\overline{T}$.  Then the kinetic term in the Lagrangian is
\beq
	{\cal L}_{\rm kin} \cong {3\over
	2\sigma}\left|\partial_\mu\phi^i\right|^2
\eeq

To justify the replacement (\ref{Ksubs}), we can derive the kinetic
term of the brane modulus by a different method, using the DBI action
for the brane \cite{DG}.  For this, we need the form of the 10D metric that
includes not only warping, but also the dependence on the K\"ahler
modulus Re($T) = e^{4u}$:
\beq
	ds^2 = {e^{-6u}\over\sqrt{h}} dx^2 + e^{2u}\sqrt{h} \,
	\tilde g^{(6)}_{ab} dy^a dy^b,
\eeq
where $\tilde g^{(6)}_{ab}$ is the metric with some fiducial volume
on the C-Y.  The factor $e^{2u}$ in the 6D part of the metric is
understandable since $e^u\sim L$; the corresponding factor $e^{-6u}$
in the 4D part is there to put the metric in the Einstein frame,
where there is no mixing of the K\"ahler modulus with the 4D graviton.
We can now compute the induced metric in the D3 world-volume,
\beq
	G_{\mu\nu} = G_{AB}\, {\partial X^A\over\partial\sigma^\mu}\,
	{\partial X^B\over\partial\sigma^\nu}
	= {e^{-6u}\over\sqrt{h}}\eta_{\mu\nu} - e^{2u}\sqrt{h}\,
	\tilde g^{(6)}_{ab} {\partial X^a\over\partial\sigma^\mu}\,
	{\partial X^b\over\partial\sigma^\nu}
\eeq
Using this in the DBI Lagrangian and Taylor-expanding in the
transverse fluctuations, we find that
\beq
	-\tau_3\sqrt{|G_{\mu\nu}|} = -a^4\tau_3 + 
	{3\tau_3\over 4\sigma}(\partial_\mu X^a)^2 + O((\partial X)^4)
\eeq
where $a$ is the value of the warp factor at the position of the
brane.  
This reveals the presence of the extra factor of $1/\sigma$ in the
brane kinetic term which we did not see previously when we were only
considering the effect of warping, and it justifies the SUGRA
description given above.  We see that the SUGRA and DBI fields are
identified through
\beq
	\phi^1 = \sqrt{3\tau_3\over 2\sigma}(X_1 + i X_2), \hbox{\it\
etc.}
\eeq

\begin{figure}[h]
\centerline{
\includegraphics[width=0.5\textwidth]{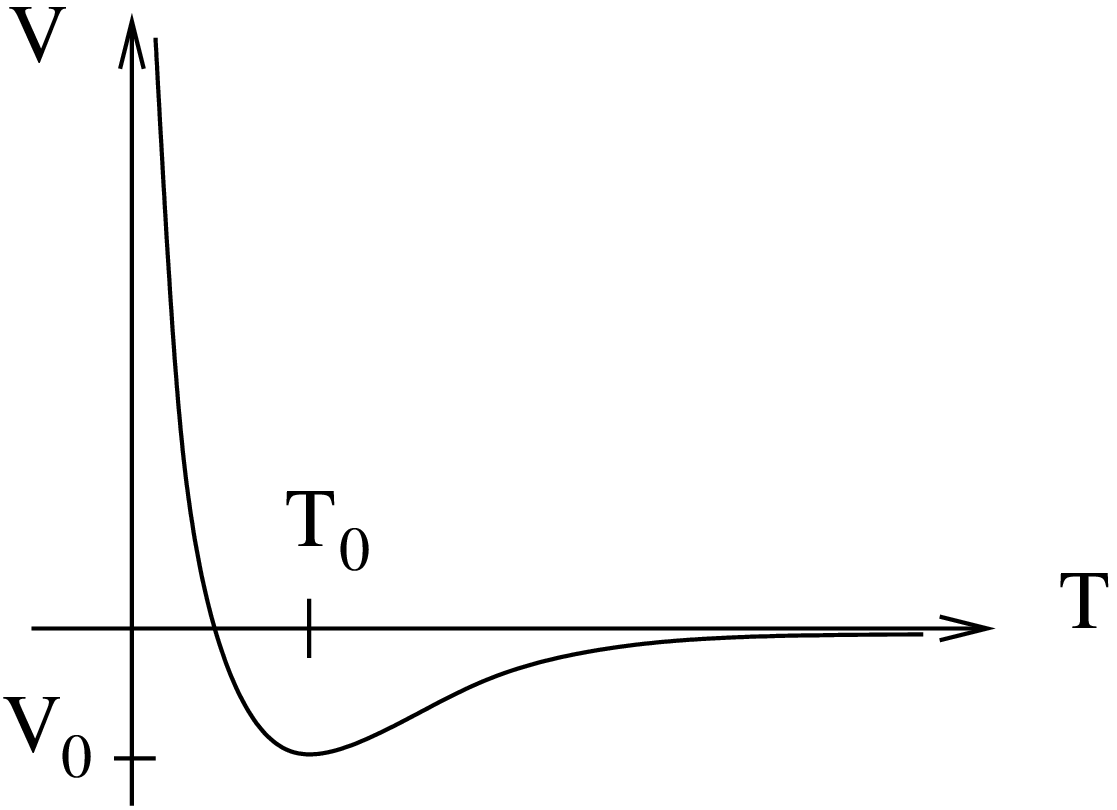}}
\caption{The KKLT potential for the K\"ahler modulus, before uplifting by 
$\overline{\hbox{D3}}$ antibrane.}
\label{kklt}
\end{figure}

With this background we can now present the $\eta$ problem.  As long
there is no potential for $T$, no new potential is
introduced for the inflaton $\phi^i$; however we know that a
stabilizing potential for $T$ must be present.  Several ways of
nonperturbatively generating such a potential were suggested by KKLT
\cite{KKLT}, which yield an extra contribution to the superpotential,
\beq
	W = W_0 + A e^{-aT}
\eeq
By itself, this new contribution indeed stabilizes $T$ at a nontrivial
value, but at the minimum the potential $V_0$ is negative, 
as shown in fig.\ \ref{kklt}, which would give
an AdS background in 4D, rather than Minkowski or de Sitter space.
To lift $V_0$ to a nonnegative value, one needs an additional,
supersymmetry-breaking contribution, which thus appears directly as
a new contribution to the potential rather than to the superpotential.
In fact, placing a $\overline{\hbox{D3}}$ antibrane in the throat does
precisely what is needed.  As we have already seen, a $\overline{\hbox{D3}}$
has positive energy, twice the warped tension,
 due to the failure of cancellation of 
the gravitational and RR potentials.  It is minimized at a nonzero
value when the  $\overline{\hbox{D3}}$ sinks to the bottom of the
throat.  The important point is that in the Einstein frame, 
taking into account the K\"ahler modulus, the $\overline{\hbox{D3}}$
contribution to the potential takes the form
\beq
	\delta V = {2 a_0^4 \tau_3\over (2{\rm Re} T)^2}
\eeq
However we have seen that when a D3 brane is in the throat, it
modifies the volume of the C-Y, and we must replace 2Re$T\to
2\sigma = T+\overline{T} - |\phi|^2$.  In addition, we get the
Coulombic interaction energy between the brane and the antibrane
so the total inflaton potential becomes
\beq
	\delta V \to {V(\phi)\over (2\sigma)^2}
	\cong {2 a_0^4 \tau_3} {1-c|\phi|^{-4}\over
	(T+\overline{T} - |\phi|^2)^2}
\eeq

The $\eta$ problem is now apparent, since even if the Coulomb
interaction is arbitrarily small ($c\to 0$), the inflaton gets a 
large mass from the term in the denominator:
\beq
	{\cal L} \sim {3\over 2\sigma}(\partial\phi)^2
	- {a_0^4\tau_3\over 2\sigma^2}\left(1 + 
	{|\phi|^2\over \sigma}\right)
\eeq
We see that the inflaton mass is of order the Hubble parameter,
\beq
	m^2 = \frac23 V= 2H^2
\eeq
(working in units where $M_p=1$), which implies that $\eta = V''/V
= m^2/(3H^2) = 2/3$, whereas inflation requires that $\eta\ll 1$. 
This is the $\eta$ problem.

\subsection{Solutions (?) to the $\eta$ problem }
A number of possible remedies to the flatness problem of
brane-antibrane inflation have been suggested.  We will see that many of
these have certain problems of their own.

\subsubsection{Superpotential corrections}
In an appendix of \cite{KKLMMT}, the possibility of having
$\phi$-dependent corrections to the superpotential was discussed,
\beq
	W = W_0 + A e^{-aT}(1 + \delta|\phi|^2 )
\eeq
which is just an ansatz for the kind of corrections which might arise
within string theory.
The inflaton mass from this modified superpotential was computed to be
\beq
	m^2 = 2H^2 
	\left(1 - {|V_{\rm AdS}|\over V_{dS}}(\beta-2\beta^2)
	\right)
\eeq
where $\beta = -\delta/a$, and $V_{\rm AdS}$ is the value of the 
potential at the AdS minimum, before uplifting with the antibrane.
By fine-tuning $\beta$, one can make
$m^2\ll H^2$ and thus get acceptable inflation.  (It was argued 
in \cite{FT2} that the tuning is only moderate, but this seems to be
contingent upon taking 3$\sigma$ rather than 1$\sigma$ error bars 
from the experimental constraints.)   
The suggestion of using superpotential corrections was 
made before any concrete calculations of the actual 
corrections had been carried out.  Since then, it has been shown 
that they have a form which drive the brane more quickly to the 
bottom of the throat rather than slowing it down; the corrections have
the wrong sign \cite{Baumann, BCDF}.  

\subsubsection{Tuning the length of the throat}

\begin{figure}[h]
\centerline{
\includegraphics[width=0.5\textwidth]{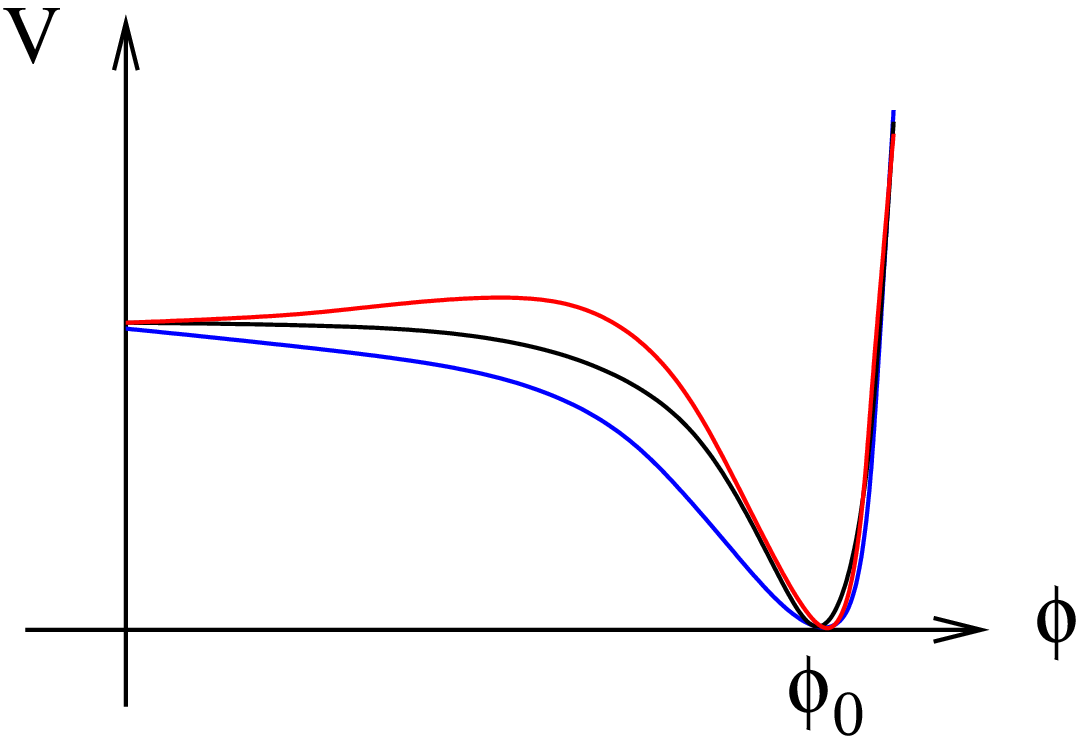}}
\caption{The brane-antibrane potential for different values of
$\phi_0$ relative to other parameters in the potential.}
\label{tuning}
\end{figure}

In KKLMMT, the impression was given that the function
$k(\phi,\bar\phi) = |\phi|^2 + O(\phi^4)$ was expanded around some
point in the C-Y which did not necessarily coincide with the bottom
of the throat; in their calculation it did not matter where this
point was because they assumed the Coulomb part of the potential was
negligible compared to the dependence via $T+\overline{T} -
k(\phi,\bar\phi)$.  However, if the origin of $\phi^i$ was assumed to
be at a position other than the bottom of the throat, thus located at
some position $\phi_0$, then there could be a competing effect between
the two sources of $\phi$-dependence, which when properly tuned could
lead to a flat potential without invoking superpotential corrections.
The full potential then had the form \cite{BCSQ}
\beq
	{\cal L} = {3T \over (2T-\phi^2)^2}(\partial\phi)^2 - 
	{2\tau_3 a_0^4\over (2T-\phi^2)^2}\left(
	1 + {a_0^4\over N (\phi-\phi_0)^4} \right)^{-1}
\eeq
where (recall that) $N = JK$ is the product of the flux quantum
numbers, and we have displayed the potential in a form which makes
sense even as $\phi\to\phi_0$ (whereas keeping only the first term of
the Taylor expansion does not).  By tuning $\phi_0$ we can adjust the
flatness of the potential in the region between $\phi=0$, assumed to
be somewhere near the top of the throat, and $\phi=\phi_0$, the
bottom.  Figure \ref{tuning} shows the effect of changing the value of
$\phi_0$ relative to other dimensionful combinations of parameters in the
potential. 

Unfortunately, the same computations \cite{Baumann, BCDF} which
clarified the nature of the superpotential corrections 
also made it clear that $\phi=0$ indeed corresponds to the bottom of
the throat, so $\phi_0$ is not a free parameter which can be tuned.
However there is a related idea that appears to be viable. 
Suppose there are two throats, related to each other by a $Z_2$
symmetry.  A brane in between them would not be able to decide 
which throat to fall into, so the potential must be flat at 
this point in the middle \cite{Trivedi}.  

\subsubsection{Multibrane inflation}
\begin{figure}[h]
\centerline{
\includegraphics[width=0.5\textwidth]{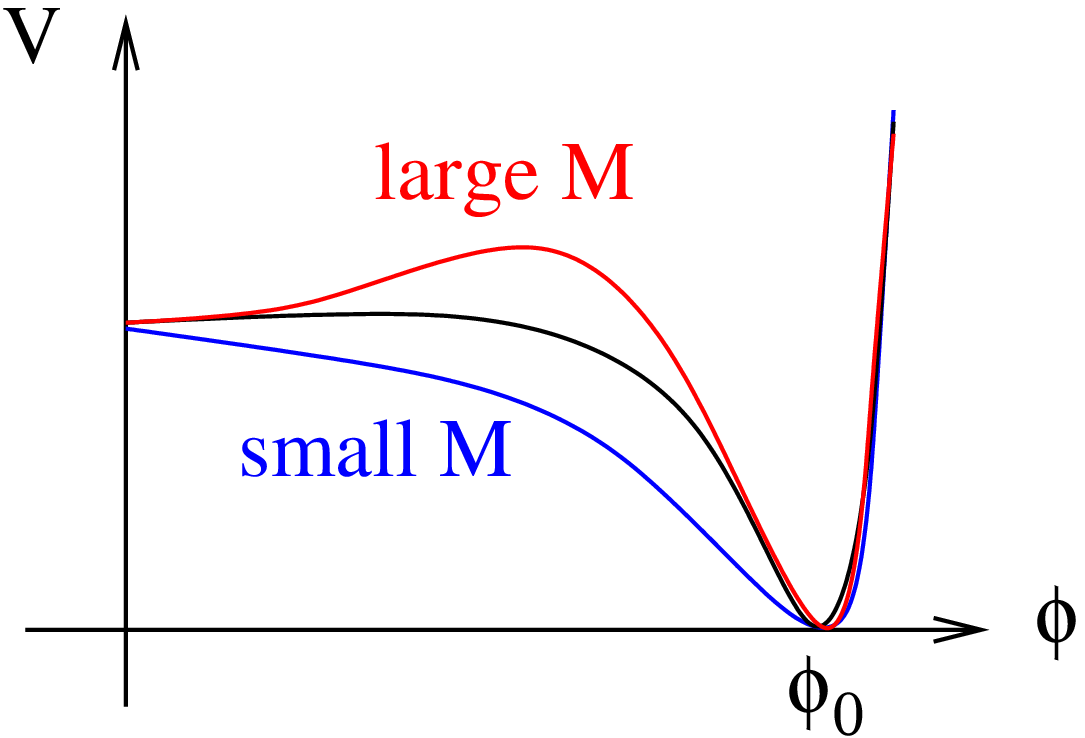}}
\caption{The brane-antibrane potential for different values of $M$,
the number of branes or antibranes in each stack.}
\label{multifig}
\end{figure}

It was also suggested that instead of using a single
D3-$\overline{\hbox{D3}}$ pair, one can use stacks of $M$ D3-branes
and $M$ $\overline{\hbox{D3}}$-branes.  If the branes in each stack 
are coincident, the Lagrangian becomes
\beq
	{\cal L} = {\cal L} = M{3T \over (2T-M\phi^2)^2}
	(\partial\phi)^2 - 
	M{2\tau_3 a_0^4\over (2T-M\phi^2)^2}\left(
	1 + M{a_0^4\over N (\phi-\phi_0)^4} \right)^{-1} 
\label{multi}
\eeq
In \cite{CS} it was shown that the shape of the potential can vary with
$M$ in such a way that for large $M$ there is a metastable minimum
in which the brane stacks remain separated, while for small $M$ the
minimum becomes a maximum (fig.\ \ref{multifig}).  For an intermediate value of $M$, the
potential is nearly flat.  The interesting feature is that $M$ can
change dynamically by tunneling of branes from the metastable minimum 
through the barrier to the stack of antibranes.  In this way one could
eventually pass through a nearly flat potential suitable for
inflation.  Unfortunately this mechanism also relied on the parameter
$\phi_0$ which was shown to be zero. 

One might have hoped that even without the possibility of tuning, having
$M$ branes in the stack could allow for assisted inflation \cite{assisted},
where the slow-roll parameters get reduced by a factor of $M$ if all
inflatons are rolling in the same way.  Unfortunately this mechanism only
works if the potential for each field by itself is independent of $M$.  In
the potential (\ref{multi}), this is not the case: the $V''/V$ 
also increases with $M$, undoing the assistance which would have come from
renormalizing the field to get a standard kinetic term.\footnote{I thank
Andrew Frey for this observation.}  On the other hand, ref.\ \cite{Becker}
finds that assisted inflation does work for multiple M5 branes moving along
the 11th dimension of M-theory.

\subsubsection{DBI inflation (D-celleration)}
References \cite{dcel,DBI} explored a different region of the
brane-antibrane parameter space and exposed a qualitatively new
possibility, in which {\it fast} roll of the D3 brane could still lead
to inflation.  In this regime, one does not expand the DBI action
in powers of $\dot\phi^2$ since it is no longer considered to be
small.  The Lagrangian has the form
\beq
	{\cal L} = -a^3(t)\left({\tau_3\over h(r)} \sqrt{1-h(r)\dot
	r^2} - V(r)\right)
\eeq
where $a(t)$ is the scale factor and $r$ is the radial position of the
D3-brane in the throat.  To be in the qualitatively new regime, one
wants the potential to be so steep that the brane is rolling nearly as
fast as the local speed limit in the throat, beyond which the argument
of the square root changes sign: $h(r)\dot r^2\sim 1$.  One finds that
this results in power-law inflation,
\beq
	a(t) \propto t^{1/\epsilon},\quad
	\epsilon \cong {M_p\over m}\, \sqrt{3 g_s\over R^4 M_s^4}
\eeq
where we recall that $R$ is the AdS curvature scale of the throat,
$M_s$ is the string scale, and $m$ is the inflaton mass.  
One sees that inflation works best
when $m$ is as large as possible, which is quite different from
inflation with a conventional kinetic term.  Even if $m\ll M_p$,
we can compenstate by taking $R\gg M_s^{-1}$ to get $\epsilon\ll 1$.

However, this takes a huge amount of flux in the original KS model.
Recall that $R^4 M_s^4 = {27\pi\over 4} g_s N$ where $N=JK$, so
\beq
	\epsilon = {M_p\over m}\sqrt{4\over 9\pi N}
\eeq
If $m$ is of order the GUT scale, then we need $J\sim K\sim 10^3$.
In fact the problem is even worse, because the COBE
normalization of the CMB power spectrum gives
\beq
	P = {1\over 4\pi^2}\, {g_s\over \epsilon^4\lambda}
\eeq
where $\lambda= R^4 M_s^4$ is identified as the `t Hooft coupling
in the context of the AdS-CFT correspondence.   The COBE normalization
implies that $\lambda\sim 10^{14}$, which requires an Euler number
for the C-Y of the same order, according to the tadpole condition
(\ref{tadpole}).  This exceeds by many orders of magnitude the largest
known example.  Ref.\ \cite{DBI} suggests some ways to overcome this
difficulty.

One of the most interesting consequences of DBI inflation is the
prediction of large nongaussianity, which is not possible in
conventional models of single-field inflation.  To see why it occurs
in the DBI model, one should consider the form of the Lagrangian
for fluctuations $\delta\phi$ of the inflaton.  The unperturbed
kinetic term has the form
\beq
	{\cal L}_{\rm kin} = -{1\over h}\sqrt{1 - h\dot\phi^2}
	\equiv {1\over h \gamma}
\eeq
where $\gamma$ is analogous to its counterpart in special relativity;
in the DBI inflation regime, we have $\gamma\gg 1$ since the field is
rolling close to its maximum speed.  
The first variation of this term is
\beqa
	\delta{\cal L}_{\rm kin} &=& {1\over h} \, {h\dot\phi\over 
	\sqrt{1-h\dot\phi^2}}\, \delta\dot\phi =
	\gamma\dot\phi\delta\dot\phi\nonumber\\
	&\cong& {\gamma\over\sqrt{h}}\,\delta\dot\phi
\eeqa
The fluctuation Lagrangian is enhanced relative to the zeroth order
Lagrangian by powers of $\gamma$.  The higher the order in
$\delta\phi$, the more powers of $\gamma$.  Nongaussian features start
appearing at order $\delta\phi^3$, through the bispectrum (3-point
function) of the fluctuations.  DBI inflation predicts that the
nonlinearity parameter, which is the conventional measure of
nongaussianity, is of order
\beq
	f_{NL} \cong 0.06 \gamma^2
\eeq
which is $\gamma^2$ times the prediction of ordinary inflation
models.  The current experimental limit is $|f_{NL}|\lsim 100$ \cite{wmap3}, 
(hence $\gamma < 40$), with a
future limit of $|f_{NL}|\lsim 5$ projected for the PLANCK experiment.
Thus DBI inflation has a chance of producing observable levels of
nongaussianity, which conventional models do not.

Moreover, DBI inflation predicts a tensor component of the CMB with
tensor-to-scalar ratio 
\beq
	r = 16{\epsilon\over \gamma}
\label{rdbi}
\eeq
so that an upper bound on nongaussianity (hence on $\gamma$) 
implies a lower bound on tensors---a kind of no-lose theorem.

The original work on DBI inflation focused on inflation far from
the tip of the throat, but recent work has pointed out that 
generically (for the KS throat background)
one tends to get the last 60 e-foldings of inflation
at the bottom the throat, and having larger-than-observed
levels of nongaussianity \cite{kec}.

\subsubsection{Shift symmetry; D3-D7 inflation}
It was suggested \cite{FT1} that there could be a symmetry of the
Lagrangian
\beqa
	\phi_i &\to& \phi_i + c_i\\
	T &\to& \bar c_i \phi_i + \frac12\sum c_i\bar c_i
\eeqa
which leaves a flat direction despite the combination $T+\overline{T}
- |\phi|^2$ in the K\"ahler potential.  In fact, it is only necessary
to preserve a remnant of this shift symmetry, $\phi_i\to \phi_i +$
Re($c_i)$  to get a flat inflaton potential; this can be achieved with
a superpotential of the form 
\beq
	W = W_0 + A e^{-a(T-\frac12\phi^2) + i\beta_i\phi_i}
\eeq
Such a symmetry requires an underlying symmetry of the
compactification manifold, namely isometries, which would exist for
toroidal compactifications.  These are not favored for realistic
model building.  

It was argued that shift symmetry can occur in models where inflation
is driven by the interaction between a D3 and a D7 brane 
\cite{keshav}.  It is interesting to note that the DBI action for
D7-branes does not get the factor of $1/\sigma$ which was the
harbinger of the $\eta$ problem for D3-$\overline{\hbox{D3}}$
inflation.  If we split the C-Y metric into two factors $\tilde
g^{(4)}$ for the extra dimensions parallel to D7, and $\tilde
g^{(2)}$ for those which are transverse, the induced metric for 
D7 gives
\beqa
	\sqrt{G_{ab}} &\cong& \sqrt{g^{(4)}\tilde g^{(2)}}\,
	\left(1- g_{(4)}^{\mu\nu}\, \tilde g^{(2)}_{ij}\,\partial_\mu S^i
	\partial_\nu S^j\right)^{1/2}\nonumber\\
	&\sim& (e^{-6u+2u})^{4/2} 
	\left(1 - \frac12 e^{6u+2u}(\partial S)^2\right)
\eeqa
and the factors of $u$ cancel out of the kinetic term 
for the two transverse fluctuations $S^i$.  

D3-D7 inflation can be pictured as a point (the D3 brane fully
localized in the extra dimensions) interacting with a 4-brane
(the 4 dimensions of the D7 which wrap the compact dimensions),
which is like a D3 smeared over 4 extra dimensions.  Thus the $1/r^4$
Coulombic potential of D3-$\overline{\hbox{D3}}$ gets integrated over
4 dimensions to become a logarithmic potential for D3-D7.  This
potential arises from SUSY-breaking effects, such as turning on
fluxes which live in the world-volume of the D7-brane.

The low-energy description of this model is a SUGRA hybrid inflation
model, with potential
\beqa
	V(S,\phi_+,\phi_-) &=& 2g^2|S|^2(|\phi_+|^2 + |\phi_-|^2)
	+
	 2g^2\left|\phi_+\phi_- -\frac12\zeta_+\right|^2\nonumber\\
	&+&\frac12 g^2\left(|\phi_+|^2-|\phi_-|^2 - \zeta_3\right)^2
\eeqa
where $S$ is the brane separation in a $T^2/Z_2$ subspace of the C-Y,
$\phi_\pm$ are the lowest modes of the strings stretched between D3
and D7, and $\zeta_3$, $\zeta_+$ are Fayet-Iliopoulos terms, which
arise due to a background Kalb-Ramond $H_3$ field strength.

$S$ is the inflaton, while the fields $\phi_\pm$ are ``waterfall''
fields, whose tachyonic instability brings about the end of inflation,
provided that $\zeta_3$ or $\zeta_+\neq 0$.  In that case
$m^2_{\phi_\pm}$ changes sign as $S\to 0$.  The instability is the
dissolution of the D3 in the D7.  At large $S$, the waterfall fields'
potentials are minimized at $\phi_\pm=0$, and the potential for $S$ is
exactly flat (as demanded by the shift symmetry).  However
SUSY-breaking generates a potential for $S$ at one-loop,
\beq
	\Delta V = {g^4\over 16\pi^2}\zeta_3^2 \ln\left|S\over S_c
	\right|^2
\eeq
as is needed to get inflation to end.  However it is still an open question
as to whether the shift symmetry gets broken more badly than this, possibly
spoiling inflation,  when the
full quantum corrections are taken into account \cite{BHK}.

\subsubsection{Racetrack and K\"ahler moduli inflation}
\begin{figure}[h]
\centerline{
\includegraphics[width=0.75\textwidth]{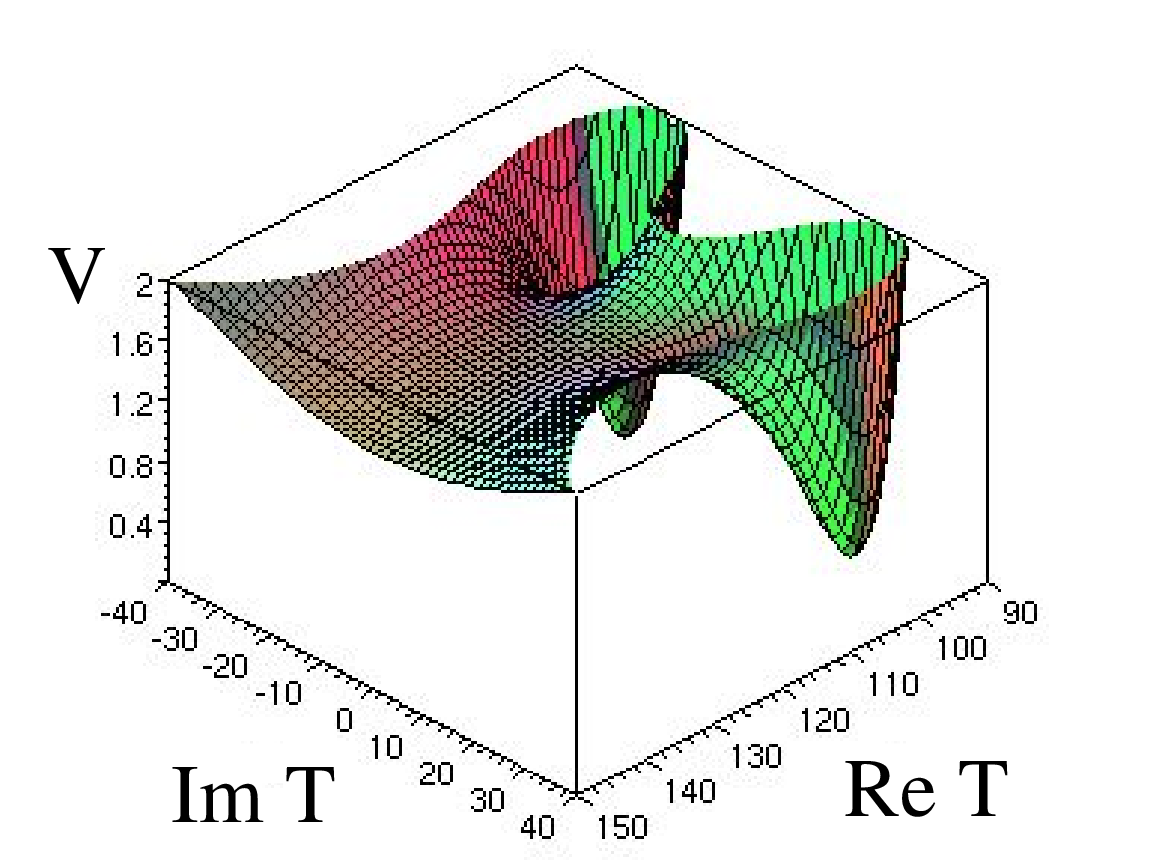}}
\caption{The potential for the K\"ahler modulus in a racetrack inflation
model.}
\label{racetrack}
\end{figure}

Another alternative is to get rid of the mobile D3-brane entirely and
instead use the compactification moduli to get inflation
\cite{rt1}.  This
requires a slightly more complicated superpotential
\beq
	W = W_0 + Ae^{-aT} + B e^{-bT}
\label{rt1}
\eeq
which for historical reasons is known as a racetrack superpotential.
It can come about from gaugino condensation in two strongly
interacting gauge groups, {\it e.g.} SU(N)$\times$SU(M), giving
$a=2\pi/N$, $b=2\pi/M$.  Another possibility is to have two different
K\"ahler moduli \cite{rt2}, with
\beq
	W = W_0 + Ae^{-aT_1} + B e^{-bT_2}
\label{rt2}
\eeq
The latter form arises from a a very specific C-Y compactification
\cite{DDF}; hence this model is rigorously based on string
theory.  By fine tuning parameters, one can find a saddle point such
that the axionic direction Im$T$ (or a linear combination of them in
the case where there are two K\"ahler moduli) is flat and gives rise
to inflation.  The potential is shown in fig.\ \ref{racetrack}.
Interestingly, both models (\ref{rt1}, \ref{rt2}) seem
to point to a spectral index of $n_s=0.95$, in agreement with the
recent WMAP measurement.  Getting a flat enough spectrum however also
seems to require uncomfortably large values for the rank of the 
gauge group(s).  A positive feature of the model is that the existence of
degenerate minima leads to inflating domain walls, an example of topological
inflation, which solves the initial value problem: it is guaranteed that
regions undergoing inflation will always exist in the universe, which are
the borderlines between regions existing in the different vacua.

The fine-tuning problem of the simplest racetrack models can be overcome
in backgrounds with a larger number of K\"ahler moduli \cite{fernando}.
In these models, the C-Y volume can naturally be quite large \cite{largeV},
and the potential can be flat since terms proportional to 
$e^{-a T_i}$ are small corrections in the large-$T_i$ regime, as long as 
there are some contributions to the potential which are not exponentially
suppressed.  Ref.\ \cite{fernando} finds that this can be realized in models
with three or more K\"ahler moduli, thus alleviating the need for fine
tuning.

\subsection{Confrontation with experiment}
Having described a few of the inflationary models which arise from
string theory, let's reconsider the experimental constraints which
can be used to test them.  

\subsubsection{Power spectrum}
The scalar power spectrum of the inflaton
is the main connection to observable physics.  It is parametrized as
\beq
	P(k) = A_s \left({k\over k_0}\right)^{n_s-1}
\eeq
where currently $A_s = 4\times 10^{-10}$ if $k_0 = 7.5 a_0 H_0$
(this is the COBE normalization \cite{lyth}) and $n_s = 0.95\pm 0.02$.
For a single-field inflationary model $P(k)$ is predicted to be
\beq
	P(k) = {1\over 50\pi^2}\, {H^4\over \frac12\dot\phi^2}
\label{pow}
\eeq
evaluated at horizon crossing, when $k/a = H$.  Using the slow-roll
equation of motion for the inflaton, $3H\dot\phi = V'$, one can show
that \cite{lyth}
\beq
	P(k) = {V\over 150\pi^2\epsilon M_p^4}
\eeq
Many of the inflation models we considered have noncanonically
normalized kinetic terms; for example supergravity models typically
have ${\cal L}_{\rm kin} \sim (\partial T/T)^2$.  It is often useful
to be able to evaluate quantities in the original field basis rather
than having to change to canonically normalized fields.  The correct
generalization of (\ref{pow}) is simply
\beq
	P(k) = {1\over 50\pi^2}\, {H^4\over {\cal L}_{\rm kin}}
\eeq

In addition to the scalar power, there is a tensor contribution to 
the temperature fluctuations, whose amplitude relative to the scalar
contribution is
\beq
	r = {A_t\over A_s} = 16\epsilon
\eeq
in terms of the slow-roll parameter $\epsilon$ (but see the 
different prediction (\ref{rdbi}) from DBI inflation).
The current limit is $r< 0.3$ \cite{wmap3}.

\subsubsection{Numerical methods}

Moreover in some models, like racetrack inflation, the inflaton is
some combination of several fields, and it is useful to have an
efficient way of numerically solving the equations of motion.  
Here is one approach.  Let
\beq
	\pi_i = {\partial{\cal L}\over \partial\dot\phi_i}
	= f_{ij}(\phi) \dot\phi_j
\eeq
which can be inverted algebraically to find $\dot\phi_i$ as a function
of the canonical momenta.  The equations of motion for the scalar
fields are
\beqa
	{d\over dt}\left(a^3 \pi_i\right) &=& a^3{\partial\over\partial
	\phi_i}({\cal L}_{\rm kin} - V)\nonumber\\
	\implies \dot\pi_i + 3H\pi_i &=& {\partial\over\partial
	\phi_i}({\cal L}_{\rm kin} - V)
\eeqa
This together with 
\beq
	\dot\phi_i = f_{ij}^{-1} \pi_j
\eeq
constitute a set of coupled first order equations which can be 
numerically integrated along with the Friedmann equation which
determines $a(t)$.  However we are usually not interested in the
actual time dependence of the solutions, but rather in the dependence
on the number of e-foldings, $N = \int H dt$.  It is therefore more 
efficient to trade $t$ for $N$ using $dN = H dt$, $H{d\over dN} =
{d\over dt}$.  This makes it unnecessary to also solve the Friedmann
equation, and one can simply transform the scalar field equations of
motion to 
\beqa
	{d\pi_i\over dN} + 3 \pi_i &=& {1\over H}
	{\partial\over\partial
	\phi_i}({\cal L}_{\rm kin} - V)\\
	{d\phi_i\over dN} &=& {1\over H} f_{ij}^{-1} \pi_j
\eeqa
where
\beq
	H = \sqrt{{\cal L}_{\rm kin} + V \over 3 M_p^2}
\eeq

\begin{figure}[h]
\centerline{
\includegraphics[width=0.5\textwidth]{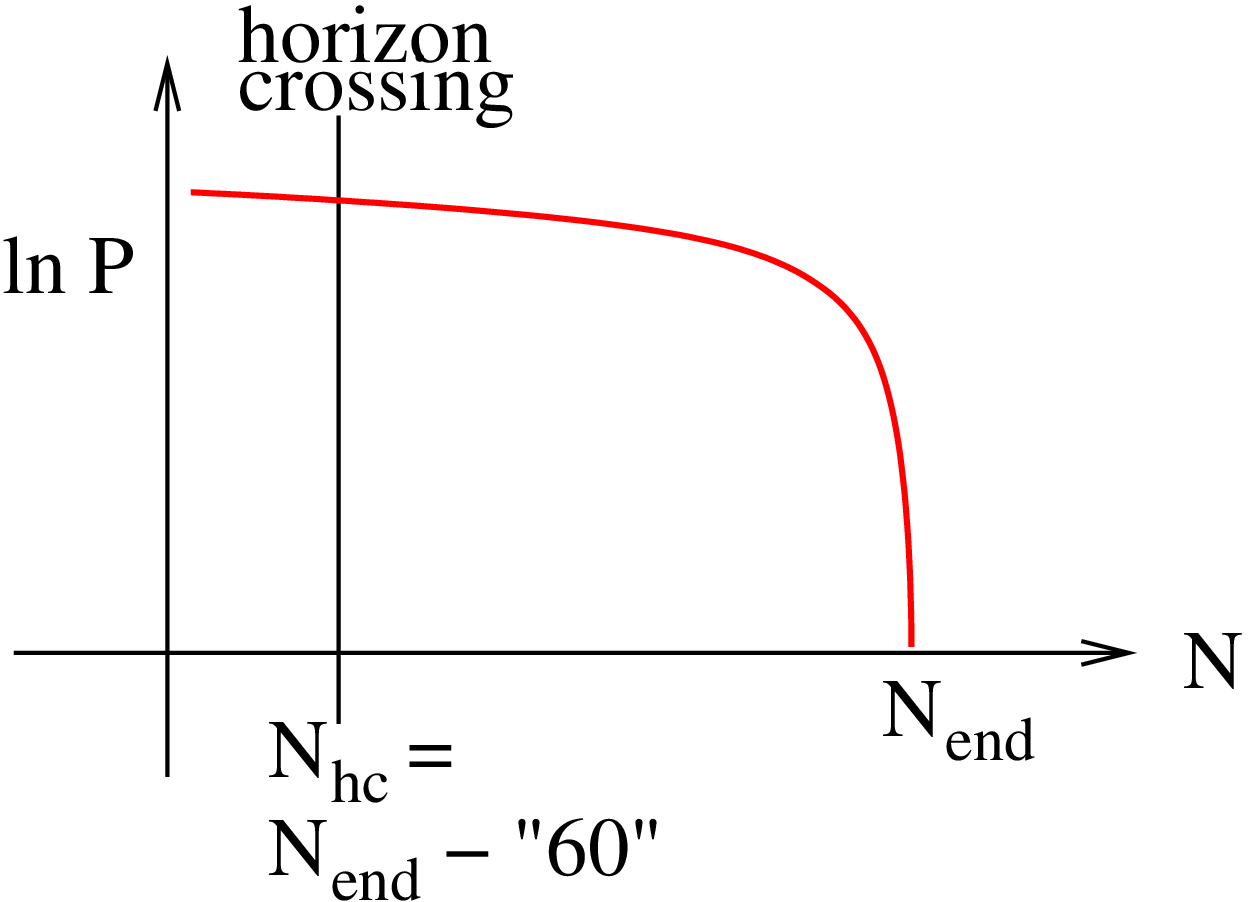}}
\caption{(a) Scalar power as a function of the number of e-foldings.
}
\label{lnP}
\end{figure}

Knowing $\phi_i(N)$ and $\pi_i(N)$, one can compute the power $P$
as a function of $N$, as shown in figure \ref{lnP}.  The tilt
at any point can be computed from
\beq
	n_s -1 = {d\ln P\over d\ln k} = {d\ln P\over dN}
	\left(1 + H^{-1}{dH\over dN}\right)^{-1}
	\cong {d\ln P\over dN}
\eeq
Here we used the horizon crossing condition $k=aH=e^N H$, hence
$\ln k = N+ \ln H$, $d\ln k = dN + d\ln H/ dN$ to change variables
from $k$ to $N$.

We are of course interested in evaluating the properties of the power
spectrum at the COBE scale, 
when perturbations roughly of the size of the present horizon
originally crossed the inflationary horizon.  This occurs at an $N$ 
given by $N=N_{hc}$, nominally 60 e-foldings before the end of
inflation.  However when we say 60, this is an upper bound, determined
by the maximum energy scale of inflation $M_{\rm inf}$
consistent with the 
experimental limit on the tensor contribution to the spectrum.  The
actual number of e-foldings since horizon crossing is given by
\beq
	N_e = 62 + \ln\left(M_{\rm inf}\over 10^{16}\hbox{\ GeV}\right)
	-\frac13\ln\left(M_{\rm inf}\over T_{rh}\right)
\eeq
where $T_{rh}$ is the reheating temperature at the end of inflation.
In the absence of detailed knowledge about reheating, one might
assume that $T_{rh}\sim M_{\rm inf}$, meaning that reheating is very
fast and efficient, or possibly that $T_{rh}\lsim 10^{10}$ GeV, since
higher reheat temperatures tend to produce too many dangerously heavy
and long-lived gravitinos \cite{ellis}.

In many models of inflation, one has the freedom by adjusting
parameters to change the overall scale of the inflationary potential
without changing its shape:
\beq
	V\to \alpha V
\eeq
In the slow-roll regime, $\alpha$ scales out of the inflaton equation
of motion, and has no effect on the number of e-foldings or the shape
of the spectrum; it only affects the normalization of $P$.  In such a
case one can always use $\alpha$ to satisfy the COBE normalization.
One can then focus attention on tuning parameters such that $n_s =
0.95$ at horizon crossing to fit the observed tilt.  

\subsubsection{New signatures---cosmic strings}

One of the great challenges for string cosmology is in producing
signals that could distinguish string models of inflation from
ordinary field theory ones.  We have seen that DBI inflation was
distinctive in that respect, since it can give large nongaussianity,
in contrast to conventional single-field models (see however
\cite{chen}).  Perhaps the the most dramatic development in this
direction has been the possibility of generating cosmologically large
superstring remnants in the sky \cite{cmp}.  To see how this can come
about, consider the effective action for the complex tachyonic 
field in the D3-$\overline{\hbox{D3}}$ system which describes the
instability.  It resembles the DBI action, but has a multiplicative
potential \cite{sen},
\beq
	S = -2\tau_3 \int d^{\,4}x\, V(|T|)\sqrt{1-|\partial T|^2}
\label{sen}
\eeq
where $V = 1/\cosh(|T|/\sqrt{2\alpha'})$.  Notice that $V$ has an
unstable maximum, indicative of a tachyon.  However 
if the branes are separated by a distance $r$, 
the potential gets modified at small $T$ such that the curvature
can become positive,
\beq
	V(|T|) \cong 1 + \left[ \left({M_s r\over 2\pi}\right)^2 -
	\frac12 \right] |M_s T|^2 + O((M_s T)^4)
\eeq
This shows that the instability turns on only when D3 and
$\overline{\hbox{D3}}$ come within a critical distance of each other.

The action (\ref{sen}) admits classical 
topologically stable string defect solutions, for example a string
oriented along the $x^3$ direction, whose solution in polar
coordinates in the $x^1$-$x^2$ plane has the form $T = T(\rho)
e^{i\theta}$.  Sen has shown that these kinds of solutions are an
exact description of D1-branes.  Ref.\ \cite{cmp} argues that
fundamental F1-strings will also be produced, as a consequence of
$S$-duality, which
exchanges $g_s\leftrightarrow g_s^{-1}$ as well as $F1\leftrightarrow
D1$.  The strings remain localized at the bottom of the KS throat
because the warp factor provides a gravitational potential barrier.

\begin{figure}[h]
\centerline{
\includegraphics[width=0.5\textwidth]{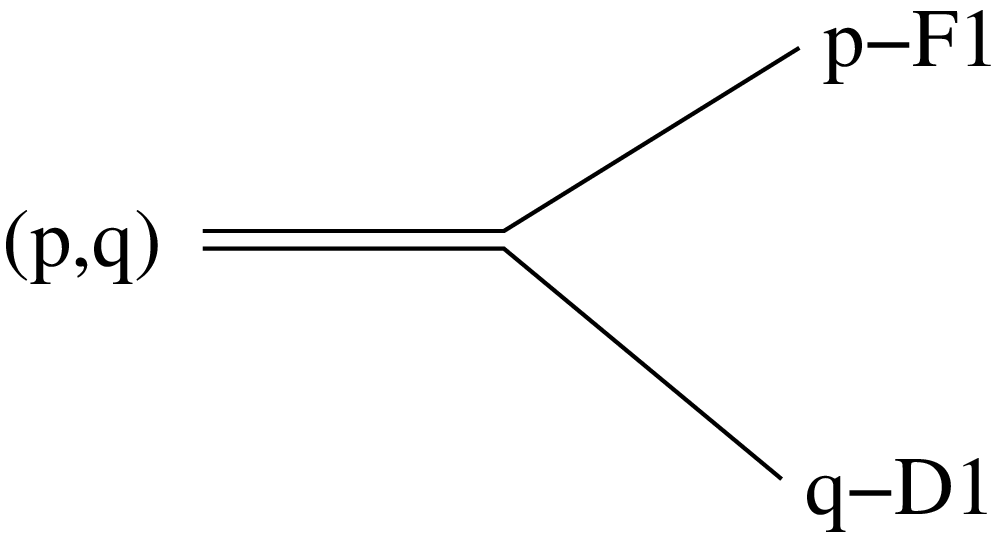}}
\caption{The junction between a $(p,q)$ string and 
a bound state of $p$ fundamental strings or $q$ D-strings.}
\label{pq}
\end{figure}

Again, field theory models can also predict relic cosmic strings,
so one would like to find ways of distingushing them from cosmic
superstrings.  One possibility is the existence of
 bound states called $(p,q)$ strings, made of
$p$ F1-strings and $p$ D1-branes, whose tension is given by
\beq
	\tau_{p,q} = {M_s^2\over 2\pi}\sqrt{ (p - C_{(0)}q)^2 + 
	q^2/g_s^2}
\eeq
where $C_{(0)}$ is the (possibly vanishing) background value 
of the RR scalar which gives rise to $F_{(1)}$. 
(For complications arising from having the strings in a warped throat,
see ref.\ \cite{FLT}.)   There exist 
three-string
junctions joining a $(p,0)$, $(0,q)$ and $(p,q)$ string, as shown in 
fig.\ \ref{pq}.  A truly
distinctive observation would be to see such a junction in the sky
using the gravitationally lensed images of background galaxies to
measure the tensions of the three strings, but we would have to be
lucky enough to be relatively close to such strings.  In the last
few years there was some initial evidence for a cosmic string lens
(CSL-1) \cite{CSL1}
, but this has been refuted by Hubble Space Telescope
observations.

\begin{figure}[h]
\centerline{
\includegraphics[width=0.75\textwidth]{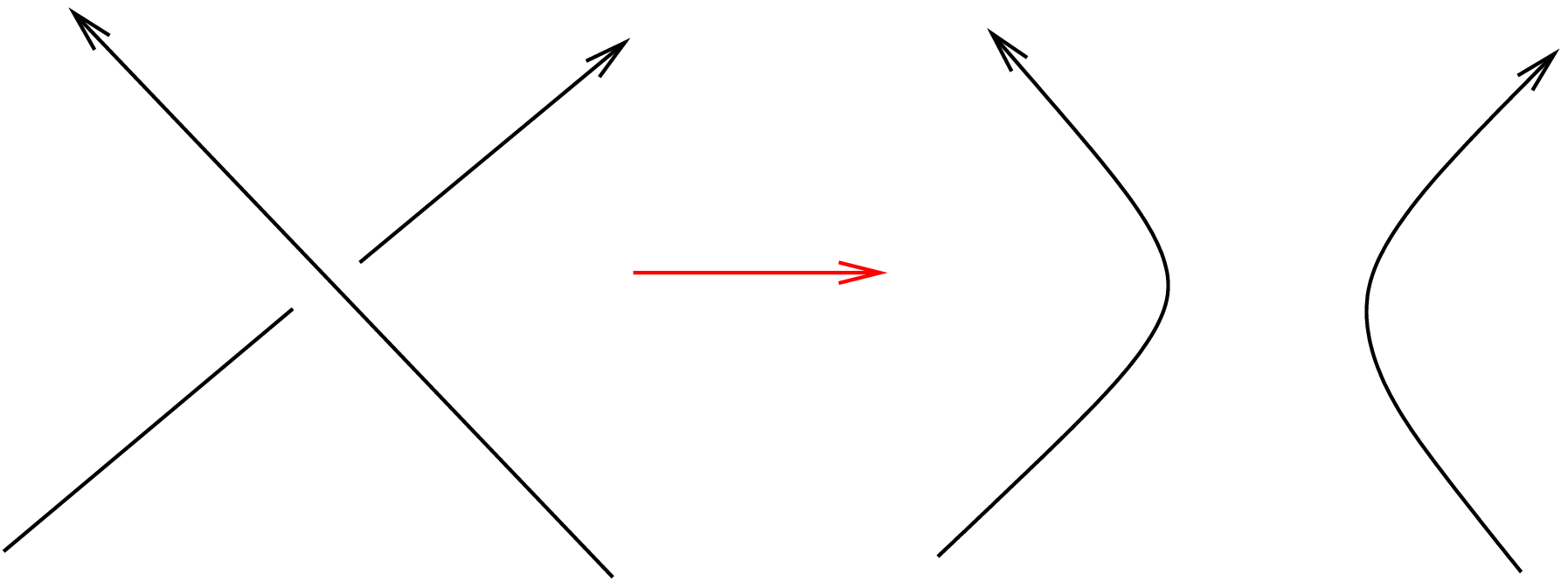}}
\caption{(a) Intercommutation of two strings after they intersect.
}
\label{intercommute}
\end{figure}

A network of relic cosmic superstrings could in principle be
distinguished from field theoretic strings by the details of their
spectrum of emitted gravity waves, which may be observable at LIGO.
All relativistic strings produce cusps during their oscillations,
which are strong emitters of gravity waves.  A major difference
between superstrings and field-theory strings is in their interaction
probabilities when two strings (or parts of the same string) cross.
Field-theory strings almost always intercommute (the ends of the
strings at the intersection point swap partners, fig.\ \ref{intercommute}), whereas
superstrings can have a smaller intercommutation probability,
$10^{-3} < P < 1$. If future LIGO observations can eventually measure
the spectrum of  stochastic gravity waves $dN/dh = A h^{-B}$, where
$h$ is the amplitude of the metric perturbation, the two parameters
$A$ and $B$ could be used to determine the intercommutation
probability $P$ and the tension of the cosmic strings, $\mu$
\cite{DV, Pol}.  In an unwarped model $\mu$ would just be the ordinary
string tension $1/\alpha'$, but this can be reduced to much lower
values in a warped throat.  The current limit on $\mu$ from gravity
waves, CMB, and especially pulsar timing is on the order of 
$G\mu < 10^{-7}.$

\subsubsection{The reheating problem}

One of the interesting consequences of brane-antibrane inflation is its new
implications for reheating at the end of inflation.  Initially (before the
warped models were introduced)  it was noticed that Sen's tachyon potential
(\ref{sen}) has no local minimum around which oscillations and conventional
reheating could occur \cite{KL}, so a new picture of how reheating might
take place was suggested, where tachyon condensation following higher
dimensional brane annihilation such as D5-$\overline{\hbox{D5}}$ would give
rise to D3-brane defects, including our own universe, which start in a hot,
excited state \cite{CF}.  However, this has the problem that the D3-brane
defects could also wrap one of the extra dimensions, appearing as a domain
wall to 3D observers, and overclosing the universe \cite{BBCS}.

With the warped models came a new twist.  While it became understood that
the tachyon condensate decays into closed strings \cite{LLM} as well as
lower dimensional branes, there was a potential problem with transferring
this energy to the visible sector of the Standard Model (SM), unless we
lived in a brane in the same throat as that where inflation was occurring. 
This was disfavored for two reasons: (1) the mass scale of the throat should
be the inflationary scale, which is typically much higher than the scale
desired for SUSY breaking to get the SM, and (2) cosmic superstrings left
over from brane-antibrane annihilation are unstable to breaking up if there
are any D3 branes left in the throat \cite{cmp}.  To overcome these
problems, one would like to localize the SM brane in a different throat from
the inflationary throat, but in this case, the warp factor acts as a
gravitational potential barrier.  It was shown that this barrier can be
overcome due to the enhanced couplings of Kaluza-Klein (KK) gravitons to the
SM brane \cite{reheating} or if there are resonant effects \cite{FT3}. 
Another mechanism for reheating is the oscillations of the SM throat which
gets deformed by the large Hubble expansion during inflation \cite{FMM}.

The KS throat can even give rise to unwanted relic KK gravitons during
reheating \cite{KY}.  The reason is that angular momentum in the $T_{1,1}$
directions can be excited within the KS throat.  If the throat was not
attached to a C-Y which broke the angular isometries, these angular momenta
would be exactly conserved, and the excited KK modes carrying them would not
be able to decay, only to annihilate, leaving some relic density of
superheavy particles to overclose the universe.  The breaking effect is
suppressed because of the fact that the KK wave functions in the throat are
exponentially small in the C-Y region where the isometries are broken.  Thus
the angular modes are unstable, but with a lifetime that is enhanced by
inverse powers of the warp factor in the throat.  It is crucial to know exactly
what power of $a_0$ arises in the decay rate for determining the
observational constraints on the model, but this seems to be quite dependent
on the detailed nature of the C-Y to which the throat is attached \cite{BCS}.

\bigskip
{\bf Acknowledgments.}  I thank N.\ Barnaby,
K.\ Dasgupta, A.\ Frey, H.\ Firouzjahi, and H.\ Stoica for many helpful comments
and answers to questions while I was preparing these lectures.
Thanks to L.\ McAllister for constructive comments on the manuscript.


\begin{thebibliography}{100}
\bibitem{eotwash}
  D.~J.~Kapner, T.~S.~Cook, E.~G.~Adelberger, J.~H.~Gundlach, B.~R.~Heckel, C.~D.~Hoyle and H.~E.~Swanson,
  ``Tests of the gravitational inverse-square law below the dark-energy length
  scale,''
  arXiv:hep-ph/0611184.

\bibitem{Susskind}
  L.~Susskind,
  ``The anthropic landscape of string theory,''
  arXiv:hep-th/0302219.

\bibitem{CDL}
  S.~R.~Coleman and F.~De Luccia,
  ``Gravitational Effects On And Of Vacuum Decay,''
  Phys.\ Rev.\ D {\bf 21}, 3305 (1980).

\bibitem{Linde}
  A.~D.~Linde,
  ``Eternal extended inflation and graceful exit from old inflation without
  Jordan-Brans-Dicke,''
  Phys.\ Lett.\ B {\bf 249}, 18 (1990).

\bibitem{WeinA}
  S.~Weinberg,
  ``Anthropic bound on the cosmological constant,''
  Phys.\ Rev.\ Lett.\  {\bf 59}, 2607 (1987).


\bibitem{WeinR}
  S.~Weinberg,
  ``The cosmological constant problem,''
  Rev.\ Mod.\ Phys.\  {\bf 61}, 1 (1989).

\bibitem{SCP}

A.~G.~Riess {\it et al.}  [Supernova Search Team Collaboration],
``Observational Evidence from Supernovae for an Accelerating Universe and a Cosmological Constant,''
Astron.\ J.\  {\bf 116}, 1009 (1998)
[arXiv:astro-ph/9805201].

S.~Perlmutter {\it et al.}  [Supernova Cosmology Project Collaboration],
``Measurements of Omega and Lambda from 42 High-Redshift Supernovae,''
Astrophys.\ J.\  {\bf 517}, 565 (1999)
[arXiv:astro-ph/9812133].


\bibitem{BP}
  R.~Bousso and J.~Polchinski,
  ``Quantization of four-form fluxes and dynamical neutralization of the
  cosmological constant,''
  JHEP {\bf 0006}, 006 (2000)
  [arXiv:hep-th/0004134].

\bibitem{BT}
  J.~D.~Brown and C.~Teitelboim,
  `Neutralization of the cosmological constant by membrane creation,''
  Nucl.\ Phys.\ B {\bf 297}, 787 (1988);
  J.~D.~Brown and C.~Teitelboim,
  ``Dynamical neutralization of the cosmological constant,''
  Phys.\ Lett.\ B {\bf 195}, 177 (1987).

\bibitem{Douglas}
  M.~R.~Douglas,
  ``Basic results in vacuum statistics,''
  Comptes Rendus Physique {\bf 5}, 965 (2004)
  [arXiv:hep-th/0409207].
 


\bibitem{dealwis}
  S.~P.~de Alwis,
  ``The scales of brane nucleation processes,''
  arXiv:hep-th/0605253.

\bibitem{Q}
  M.~Tegmark and M.~J.~Rees,
  ``Why is the CMB fluctuation level $10^{-5}$?,''
  Astrophys.\ J.\  {\bf 499}, 526 (1998)
  [arXiv:astro-ph/9709058].

  M.~L.~Graesser, S.~D.~H.~Hsu, A.~Jenkins and M.~B.~Wise,
  ``Anthropic distribution for cosmological constant and primordial density
  perturbations,''
  Phys.\ Lett.\ B {\bf 600}, 15 (2004)
  [arXiv:hep-th/0407174].

\bibitem{ADK}
N.~Arkani-Hamed, S.~Dimopoulos and S.~Kachru,
  ``Predictive landscapes and new physics at a TeV,''
  arXiv:hep-th/0501082.

\bibitem{anthropic}


S.~A.~Abel, C.~S.~Chu, J.~Jaeckel and V.~V.~Khoze,
  ``SUSY breaking by a metastable ground state: Why the early universe
  preferred the non-supersymmetric vacuum,''
  arXiv:hep-th/0610334.

N.~J.~Craig, P.~J.~Fox and J.~G.~Wacker,
  ``Reheating metastable O'Raifeartaigh models,''
  arXiv:hep-th/0611006.

W.~Fischler, V.~Kaplunovsky, C.~Krishnan, L.~Mannelli and M.~Torres,
  ``Meta-stable supersymmetry breaking in a cooling universe,''
  arXiv:hep-th/0611018.


S.~A.~Abel, J.~Jaeckel and V.~V.~Khoze,
  ``Why the early universe preferred the non-supersymmetric vacuum. II,''
  arXiv:hep-th/0611130.


L.~Pogosian and A.~Vilenkin,
  ``Anthropic predictions for vacuum energy and neutrino masses in the light of
  WMAP-3,''
  arXiv:astro-ph/0611573.

 D.~Schwartz-Perlov,
  ``Probabilities in the Arkani-Hamed-Dimopolous-Kachru landscape,''
  arXiv:hep-th/0611237.

\bibitem{DGT}
  M.~Dine, E.~Gorbatov and S.~D.~Thomas,
  ``Low energy supersymmetry from the landscape,''
  arXiv:hep-th/0407043.

\bibitem{AHD}
  N.~Arkani-Hamed and S.~Dimopoulos,
  ``Supersymmetric unification without low energy supersymmetry and  signatures
  for fine-tuning at the LHC,''
  JHEP {\bf 0506}, 073 (2005)
  [arXiv:hep-th/0405159].

\bibitem{split}
  G.~F.~Giudice and A.~Romanino,
  ``Split supersymmetry,''
  Nucl.\ Phys.\ B {\bf 699}, 65 (2004)
  [Erratum-ibid.\ B {\bf 706}, 65 (2005)]
  [arXiv:hep-ph/0406088].


\bibitem{dcel}
  E.~Silverstein and D.~Tong,
  ``Scalar speed limits and cosmology: Acceleration from D-cceleration,''
  Phys.\ Rev.\ D {\bf 70}, 103505 (2004)
  [arXiv:hep-th/0310221].

\bibitem{DBI}
  M.~Alishahiha, E.~Silverstein and D.~Tong,
  ``DBI in the sky,''
  Phys.\ Rev.\ D {\bf 70}, 123505 (2004)
  [arXiv:hep-th/0404084].


\bibitem{KS}
 I.~R.~Klebanov and M.~J.~Strassler,
  ``Supergravity and a confining gauge theory: Duality cascades and
  $\chi$SB-resolution of naked singularities,''
  JHEP {\bf 0008}, 052 (2000)
  [arXiv:hep-th/0007191].

\bibitem{GKP}
 S.~B.~Giddings, S.~Kachru and J.~Polchinski,
  ``Hierarchies from fluxes in string compactifications,''
  Phys.\ Rev.\ D {\bf 66}, 106006 (2002)
  [arXiv:hep-th/0105097].

\bibitem{polbook}
  J.~Polchinski,
  ``String theory. Vol. 2: Superstring theory and beyond,''

\bibitem{BS95}
  T.~Banks and L.~Susskind,
  ``Brane - Antibrane Forces,''
  arXiv:hep-th/9511194.

\bibitem{DT}
  G.~R.~Dvali and S.~H.~H.~Tye,
  ``Brane inflation,''
  Phys.\ Lett.\ B {\bf 450}, 72 (1999)
  [arXiv:hep-ph/9812483].

\bibitem{Burgess}
  C.~P.~Burgess, M.~Majumdar, D.~Nolte, F.~Quevedo, G.~Rajesh and R.~J.~Zhang,
  ``The inflationary brane-antibrane universe,''
  JHEP {\bf 0107}, 047 (2001)
  [arXiv:hep-th/0105204].

\bibitem{wmap3}
  D.~N.~Spergel {\it et al.},
  ``Wilkinson Microwave Anisotropy Probe (WMAP) three year results:
  Implications for cosmology,''
  arXiv:astro-ph/0603449.

\bibitem{RS}
  L.~Randall and R.~Sundrum,
  ``A large mass hierarchy from a small extra dimension,''
  Phys.\ Rev.\ Lett.\  {\bf 83}, 3370 (1999)
  [arXiv:hep-ph/9905221];
  ``An alternative to compactification,''
  Phys.\ Rev.\ Lett.\  {\bf 83}, 4690 (1999)
  [arXiv:hep-th/9906064].


\bibitem{KKLMMT}
S.~Kachru, R.~Kallosh, A.~Linde, J.~M.~Maldacena, L.~McAllister and S.~P.~Trivedi,
  ``Towards inflation in string theory,''
  JCAP {\bf 0310}, 013 (2003)
  [arXiv:hep-th/0308055].

\bibitem{HKN}
  P.~Candelas and X.~C.~de la Ossa,
  ``Comments on conifolds,''
  Nucl.\ Phys.\ B {\bf 342}, 246 (1990).

  K.~Higashijima, T.~Kimura and M.~Nitta,
   ``Supersymmetric nonlinear sigma models on Ricci-flat Kaehler manifolds  with
  O(N) symmetry,''
  Phys.\ Lett.\ B {\bf 515}, 421 (2001)
  [hep-th/0104184].


\bibitem{DG}
  O.~DeWolfe and S.~B.~Giddings,
  ``Scales and hierarchies in warped compactifications and brane worlds,''
  Phys.\ Rev.\ D {\bf 67}, 066008 (2003)
  [hep-th/0208123].

\bibitem{KKLT}
 S.~Kachru, R.~Kallosh, A.~Linde and S.~P.~Trivedi,
  ``De Sitter vacua in string theory,''
  Phys.\ Rev.\ D {\bf 68}, 046005 (2003)
  [arXiv:hep-th/0301240].


\bibitem{FT2}
  H.~Firouzjahi and S.~H.~Tye,
  ``Brane inflation and cosmic string tension in superstring theory,''
  JCAP {\bf 0503}, 009 (2005)
  [arXiv:hep-th/0501099].

\bibitem{Baumann}
  D.~Baumann, A.~Dymarsky, I.~R.~Klebanov, J.~Maldacena, L.~McAllister and A.~Murugan,
  ``On D3-brane potentials in compactifications with fluxes and wrapped
  D-branes,''
  JHEP {\bf 0611}, 031 (2006)
  [arXiv:hep-th/0607050].

\bibitem{BCDF}
  C.~P.~Burgess, J.~M.~Cline, K.~Dasgupta and H.~Firouzjahi,
  ``Uplifting and inflation with D3 branes,''
  arXiv:hep-th/0610320.

\bibitem{BCSQ}
  C.~P.~Burgess, J.~M.~Cline, H.~Stoica and F.~Quevedo,
  ``Inflation in realistic D-brane models,''
  JHEP {\bf 0409}, 033 (2004)
  [arXiv:hep-th/0403119].

\bibitem{Trivedi}
  N.~Iizuka and S.~P.~Trivedi,
  ``An inflationary model in string theory,''
  Phys.\ Rev.\ D {\bf 70}, 043519 (2004)
  [arXiv:hep-th/0403203].

\bibitem{CS}
  J.~M.~Cline and H.~Stoica,
  ``Multibrane inflation and dynamical flattening of the inflaton  potential,''
  Phys.\ Rev.\ D {\bf 72}, 126004 (2005)
  [arXiv:hep-th/0508029].

\bibitem{assisted}
  A.~R.~Liddle, A.~Mazumdar and F.~E.~Schunck,
  ``Assisted inflation,''
  Phys.\ Rev.\ D {\bf 58}, 061301 (1998)
  [arXiv:astro-ph/9804177].

\bibitem{Becker}
  K.~Becker, M.~Becker and A.~Krause,
  ``M-theory inflation from multi M5-brane dynamics,''
  Nucl.\ Phys.\ B {\bf 715}, 349 (2005)
  [arXiv:hep-th/0501130].

\bibitem{kec}
S.~Kecskemeti, J.~Maiden, G.~Shiu and B.~Underwood,
  ``DBI inflation in the tip region of a warped throat,''
  JHEP {\bf 0609}, 076 (2006)
  [arXiv:hep-th/0605189].


\bibitem{FT1}
 H.~Firouzjahi and S.~H.~H.~Tye,
  ``Closer towards inflation in string theory,''
  Phys.\ Lett.\ B {\bf 584}, 147 (2004)
  [arXiv:hep-th/0312020].

\bibitem{keshav}
  K.~Dasgupta, C.~Herdeiro, S.~Hirano and R.~Kallosh,
  ``D3/D7 inflationary model and M-theory,''
  Phys.\ Rev.\ D {\bf 65}, 126002 (2002)
  [arXiv:hep-th/0203019].

J.~P.~Hsu and R.~Kallosh,
  ``Volume stabilization and the origin of the inflaton shift symmetry in
  string theory,''
  JHEP {\bf 0404}, 042 (2004)
  [arXiv:hep-th/0402047].

\bibitem{BHK}
  M.~Berg, M.~Haack and B.~Kors,
  ``Loop corrections to volume moduli and inflation in string theory,''
  Phys.\ Rev.\ D {\bf 71}, 026005 (2005)
  [arXiv:hep-th/0404087].
 ``String loop corrections to Kaehler potentials in orientifolds,''
  JHEP {\bf 0511}, 030 (2005)
  [arXiv:hep-th/0508043];
  ``On volume stabilization by quantum corrections,''
  Phys.\ Rev.\ Lett.\  {\bf 96}, 021601 (2006)
  [arXiv:hep-th/0508171].

\bibitem{rt1}
  J.~J.~Blanco-Pillado {\it et al.},
  ``Racetrack inflation,''
  JHEP {\bf 0411}, 063 (2004)
  [arXiv:hep-th/0406230].


\bibitem{rt2}
  ``Inflating in a better racetrack,''
  JHEP {\bf 0609}, 002 (2006)
  [arXiv:hep-th/0603129].

\bibitem{DDF}
  F.~Denef, M.~R.~Douglas and B.~Florea,
  ``Building a better racetrack,''
  JHEP {\bf 0406}, 034 (2004)
  [arXiv:hep-th/0404257].




\bibitem{fernando}
  J.~P.~Conlon and F.~Quevedo,
  ``Kaehler moduli inflation,''
  JHEP {\bf 0601}, 146 (2006)
  [arXiv:hep-th/0509012].

\bibitem{largeV}
  V.~Balasubramanian and P.~Berglund,
  ``Stringy corrections to Kaehler potentials, SUSY breaking, and the
  cosmological constant problem,''
  JHEP {\bf 0411}, 085 (2004)
  [arXiv:hep-th/0408054];

  V.~Balasubramanian, P.~Berglund, J.~P.~Conlon and F.~Quevedo,
  ``Systematics of moduli stabilisation in Calabi-Yau flux
  compactifications,''
  JHEP {\bf 0503}, 007 (2005)
  [arXiv:hep-th/0502058].


  J.~P.~Conlon, F.~Quevedo and K.~Suruliz,
  ``Large-volume flux compactifications: Moduli spectrum and D3/D7 soft
  supersymmetry breaking,''
  JHEP {\bf 0508}, 007 (2005)
  [arXiv:hep-th/0505076].

\bibitem{lyth}
  D.~H.~Lyth and A.~Riotto,
  ``Particle physics models of inflation and the cosmological density
  perturbation,''
  Phys.\ Rept.\  {\bf 314}, 1 (1999)
  [arXiv:hep-ph/9807278].

\bibitem{ellis}
  J.~R.~Ellis, J.~E.~Kim and D.~V.~Nanopoulos,
  ``Cosmological Gravitino Regeneration And Decay,''
  Phys.\ Lett.\ B {\bf 145}, 181 (1984).

\bibitem{chen}
  X.~Chen, M.~x.~Huang, S.~Kachru and G.~Shiu,
  ``Observational signatures and non-Gaussianities of general single field
  inflation,''
  arXiv:hep-th/0605045.

\bibitem{cmp}
  E.~J.~Copeland, R.~C.~Myers and J.~Polchinski,
  ``Cosmic F- and D-strings,''
  JHEP {\bf 0406}, 013 (2004)
  [arXiv:hep-th/0312067];

  ``Cosmic superstrings II,''
  Comptes Rendus Physique {\bf 5}, 1021 (2004).

\bibitem{sen}
A.~Sen,
``Tachyon matter,''
JHEP {\bf 0207}, 065 (2002)
[arXiv:hep-th/0203265].

A.~Sen,
``Field theory of tachyon matter,''
Mod.\ Phys.\ Lett.\ A {\bf 17}, 1797 (2002)
[arXiv:hep-th/0204143].

A.~Sen,
``Dirac-Born-Infeld action on the tachyon kink and vortex,''
Phys.\ Rev.\ D {\bf 68}, 066008 (2003)
[arXiv:hep-th/0303057].

\bibitem{FLT}
  H.~Firouzjahi, L.~Leblond and S.~H.~Henry Tye,
  ``The (p,q) string tension in a warped deformed conifold,''
  JHEP {\bf 0605}, 047 (2006)
  [arXiv:hep-th/0603161].

\bibitem{CSL1}
  M.~Sazhin {\it et al.},
  ``CSL-1: a chance projection effect or serendipitous discovery of a
  gravitational lens induced by a cosmic string?,''
  Mon.\ Not.\ Roy.\ Astron.\ Soc.\  {\bf 343}, 353 (2003)
  [arXiv:astro-ph/0302547].

 M.~V.~Sazhin, M.~Capaccioli, G.~Longo, M.~Paolillo and O.~S.~Khovanskaya,
  ``The true nature of CSL-1,''
  arXiv:astro-ph/0601494.

\bibitem{DV}
  T.~Damour and A.~Vilenkin,
  ``Gravitational radiation from cosmic (super)strings: Bursts, stochastic
  background, and observational windows,''
  Phys.\ Rev.\ D {\bf 71}, 063510 (2005)
  [arXiv:hep-th/0410222].

\bibitem{Pol}
  J.~Polchinski,
  ``Introduction to cosmic F- and D-strings,''
  arXiv:hep-th/0412244.

\bibitem{KL}
  L.~Kofman and A.~Linde,
  ``Problems with tachyon inflation,''
  JHEP {\bf 0207}, 004 (2002)
  [arXiv:hep-th/0205121].

\bibitem{CF}

  J.~M.~Cline, H.~Firouzjahi and P.~Martineau,
  ``Reheating from tachyon condensation,''
  JHEP {\bf 0211}, 041 (2002)
  [arXiv:hep-th/0207156].

  J.~M.~Cline and H.~Firouzjahi,
  ``Real-time D-brane condensation,''
  Phys.\ Lett.\ B {\bf 564}, 255 (2003)
  [arXiv:hep-th/0301101].

 N.~Barnaby and J.~M.~Cline,
  ``Creating the universe from brane-antibrane annihilation,''
  Phys.\ Rev.\ D {\bf 70}, 023506 (2004)
  [arXiv:hep-th/0403223].

\bibitem{BBCS}
  N.~Barnaby, A.~Berndsen, J.~M.~Cline and H.~Stoica,
  ``Overproduction of cosmic superstrings,''
  JHEP {\bf 0506}, 075 (2005)
  [arXiv:hep-th/0412095].


\bibitem{LLM}
  N.~Lambert, H.~Liu and J.~M.~Maldacena,
  ``Closed strings from decaying D-branes,''
  arXiv:hep-th/0303139.

\bibitem{reheating}

  N.~Barnaby, C.~P.~Burgess and J.~M.~Cline,
  ``Warped reheating in brane-antibrane inflation,''
  JCAP {\bf 0504}, 007 (2005)
  [arXiv:hep-th/0412040].

  D.~Chialva, G.~Shiu and B.~Underwood,
  ``Warped reheating in multi-throat brane inflation,''
  JHEP {\bf 0601}, 014 (2006)
  [arXiv:hep-th/0508229].

\bibitem{FT3}

  H.~Firouzjahi and S.~H.~Tye,
  ``The shape of gravity in a warped deformed conifold,''
  JHEP {\bf 0601}, 136 (2006)
  [arXiv:hep-th/0512076].

\bibitem{FMM}
  A.~R.~Frey, A.~Mazumdar and R.~Myers,
  ``Stringy effects during inflation and reheating,''
  Phys.\ Rev.\ D {\bf 73}, 026003 (2006)
  [arXiv:hep-th/0508139].

\bibitem{KY}
 L.~Kofman and P.~Yi,
  ``Reheating the universe after string theory inflation,''
  Phys.\ Rev.\ D {\bf 72}, 106001 (2005)
  [arXiv:hep-th/0507257].

\bibitem{BCS}
A.\ Berndsen, J.M.\ Cline and H.\ Stoica, work in progress 


\end{thebibliography}
\end{document}